\documentclass{article}
\usepackage{arxiv}
\usepackage{amssymb}
\usepackage{amsfonts}
\usepackage{amsthm}
\usepackage[notextcomp]{stix}
\usepackage{graphicx}
\usepackage{textcomp}
\usepackage{units}
\usepackage{colortbl}
\usepackage{xcolor}
\usepackage[font=small]{caption}
\usepackage{tikz}
\usepackage{pgfplots}
\usepackage[T1]{fontenc}
\usepackage{amsmath}
\usepackage{array}                
\usepackage{url}
\usepackage{multirow} 
\usepackage{subfigure}
\usepackage{bm}
\usepackage{mathtools}
\usepackage{comment}
\usepackage{multicol}
\usepackage{float}
 \usepackage{balance}
\usepackage{xspace}
\usepackage{algorithm}
\usepackage{algorithmicx}
\usepackage{algpseudocode}
\usepackage{booktabs}
\usepackage{soul}
\usepackage{enumitem}
\usepackage{tikzpeople}
\usepackage{cite}
\usepackage{varwidth}
\usepackage[british]{babel}
\usepackage[hidelinks]{hyperref}
\usetikzlibrary{patterns}
\usetikzlibrary{positioning}
\usetikzlibrary{shapes.geometric, arrows}
\pgfplotsset{compat=1.14}

\newcommand{\PreserveBackslash}[1]{\let\temp=\\#1\let\\=\temp}
\newcolumntype{C}[1]{>{\PreserveBackslash\centering}p{#1}}
\newcolumntype{R}[1]{>{\PreserveBackslash\raggedleft}p{#1}}
\newcolumntype{L}[1]{>{\PreserveBackslash\raggedright}p{#1}}

\newcommand{\sys}{{\sc \small Bedrock}\xspace}

\newcommand{\flb}{FLB\xspace}
\newcommand{\ftb}{FTB\xspace}

\newcommand{\req}{{\sf \small request}\xspace}

\newcommand{\one}{{\sf \small pre-prepare}\xspace}
\newcommand{\two}{{\sf \small prepare}\xspace}

\newcommand{\three}{{\sf \small commit}\xspace}
\newcommand{\reply}{{\sf \small reply}\xspace}

\newcommand{\validp}[2][black,fill=green]{\tikz[baseline=-0.5ex]\draw[#1,radius=#2] (0,0) circle ;}
\newcommand{\invalidp}[2][black,fill=red]{\tikz[baseline=-0.5ex]\draw[#1,radius=#2] (0,0) circle ;}%

\newtheoremstyle{boldremark}
{0.7em} 
{\dimexpr\topsep/2\relax} 
{}          
{}          
{\bfseries} 
{:}         
{0.2em}      
{}          

\theoremstyle{boldremark}
\newtheorem{tr}{{\bf Design Choice}}
\newtheorem{infra}{{\bf E}}
\newtheorem{struct}{{\bf P}}
\newtheorem{feature}{{\bf Q}}
\newtheorem{optim}{{\bf O}}

\title{The Bedrock of Byzantine Fault Tolerance: A Unified Platform for BFT Protocol Design and Implementation}

 \author{
 Mohammad Javad Amiri$^1$ \quad Chenyuan Wu$^1$ \quad Divyakant Agrawal$^2$\\
 {\bf Amr El Abbadi$^2$ \quad Boon Thau Loo$^1$ \quad Mohammad Sadoghi$^3$}\\
$^1$Department of Computer and Information Science, University of Pennsylvania\\
$^2$Department of Computer Science, University of California Santa Barbara\\
$^3$Department of Computer Science, University of California Davis\\
$^1$\{mjamiri, wucy, boonloo\}@seas.upenn.edu, $^2$\{agrawal, amr\}@cs.ucsb.edu, $^3$ msadoghi@ucdavis.edu
\vspace{2em}
}

\begin{document}

\maketitle

\begin{abstract}
Byzantine Fault-Tolerant (BFT) protocols have recently been extensively used by
decentralized data management systems with non-trustworthy infrastructures,
e.g., permissioned blockchains.
BFT protocols cover a broad spectrum of
design dimensions from infrastructure settings such as the communication topology,
to more technical features such as commitment strategy
and even fundamental social choice properties like order-fairness.
The proliferation of different BFT protocols has rendered it difficult to navigate the BFT landscape,
let alone determine the protocol that best meets application needs.
This paper presents {\em \sys}, a unified platform for BFT protocols design, analysis, implementation, and experiments.
\sys proposes a design space consisting of a set of design choices capturing the trade-offs
between different design space dimensions and 
providing fundamentally new insights into the strengths and weaknesses of BFT protocols.
\sys enables users to analyze and experiment with BFT protocols within the space of plausible choices,
evolve current protocols to design new ones, and even uncover previously unknown protocols.
Our experimental results demonstrate the capability of \sys to uniformly evaluate BFT protocols in new ways that were not possible before due to the diverse assumptions made by these protocols.
The results validate \sys's ability to analyze and derive BFT protocols.
\end{abstract}
\section{Introduction}\label{sec:intro}

Large-scale data management systems rely on fault-tolerant protocols
to provide robustness and high availability \cite{birman1985implementing,moser1999eternal,corbett2013spanner,decandia2007dynamo,bronson2013tao,kallman2008h,baker2011megastore}.
While cloud systems, e.g.,
Google's Spanner~\cite{corbett2013spanner}, Amazon's
Dynamo \cite{decandia2007dynamo}, and Facebook's Tao \cite{bronson2013tao},
rely on crash fault-tolerant protocols, e.g., Paxos \cite{lamport2001paxos},
to establish consensus,
a Byzantine fault-tolerant (BFT) protocol is a key ingredient in
decentralized data management systems with non-trustworthy infrastructures.
In particular, a BFT protocol is the core component of the most recent large-scale data management system, permissioned blockchains
\cite{baudet2019state,amiri2019caper,morgan2016quorum,amiri2021sharper,gupta2020resilientdb,kwon2014tendermint,androulaki2018hyperledger,gorenflo2019fastfabric,gorenflo2020xox,sharma2019blurring,ruan2020transactional,iroha,corda,greenspan2015multichain,amiri2022qanaat,chain,qi2021bidl,buchnik13fireledger}.
BFT protocols have also been used in permissionless blockchains~\cite{kokoris2018omniledger,zamani2018rapidchain,luu2016secure,kogias2016enhancing,brown2016corda},
distributed file systems~\cite{castro2002practical,adya2002farsite, clement2009upright}, 
locking service~\cite{clement2009making},
firewalls~\cite{bessani2008crutial,garcia2016sieveq,sousa2009highly,roeder2010proactive,garcia2013intrusion,yin2003separating},
certificate authority systems \cite{zhou2002coca},
SCADA systems \cite{zhou2002secure,babay2019deploying,nogueira2018challenges,kirsch2013survivable},
key-value datastores \cite{goodson2004efficient,hendricks2007low,bessani2013depsky,roeder2010proactive,dobre2013powerstore}, and
key management \cite{malkhi1998secure}.

BFT protocols use the State Machine Replication (SMR) technique \cite{lamport1978time,schneider1990implementing}
to ensure that non-faulty replicas execute client requests in the same order
despite the concurrent failure of $f$ Byzantine replicas.
BFT SMR protocols are different along several dimensions, including
the number of replicas,
processing strategy (i.e., optimistic, pessimistic, or robust), supporting load balancing, etc.
While dependencies and trade-offs among these dimensions lead to several design choices,
there is currently no unifying tool that provides the foundations for studying and analyzing
BFT protocols' design dimensions and their trade-offs.
We envision that such a unifying foundation will be based on
an in-depth understanding of existing BFT protocols and trade-offs among dimensions;
and include an API that allows protocol designers to choose among several dimensions,
and find a protocol that best fits the characteristics of their applications.

This paper presents {\em \sys}, a unified platform that enables us to
design, analyze, discover, implement, and experiment with partially asynchronous SMR BFT protocols
within the design space of possible variants.
It provides an API that enables BFT protocol designers to
analyze and experiment with BFT protocols and their trade-offs and even derive new protocols.
Application developers also can query their required BFT protocol characteristics where
the platform responds with a list of candidate BFT protocols that match the given query and enables them
to choose the BFT protocol that best fits the characteristics of their applications.

\sys presents a design space to characterize BFT protocols based on different dimensions
that capture the environmental settings, protocol structure, QoS features, and performance optimizations.
Each protocol is a plausible point in the design space.
Within the design space, \sys defines a set of design choices that
demonstrate trade-offs between different dimensions.
For example, the communication complexity can be reduced by increasing the number of communication phases or
the number of phases can be reduced by adding more replicas.
Each design choice expresses a {\em one-to-one transformation function}
to map a plausible input point (i.e., a BFT protocol) to a plausible output point (i.e., another BFT protocol) in the design space.

\sys can be used to analyze and navigate the evergrowing BFT landscape to principally compare and differentiate among BFT protocols.
On one hand,
\sys design space and its design choices give fundamentally new insights into the strengths and weaknesses of existing BFT protocols.
On the other hand, \sys enables new ways to experimentally evaluate BFT protocols by providing a unified
deployment and experimentation environment, resulting in the ability to compare different protocols proposed in diverse settings and contexts under one unified framework.

The \sys tool has several practical uses:

\begin{itemize}
\item {\bf Analyze and experiment with existing BFT protocols.}
First, \sys supports within one unified platform a wide range of existing BFT protocols, e.g., 
PBFT \cite{castro1999practical}, SBFT \cite{gueta2019sbft}, HotStuff \cite{yin2019hotstuff}, Kauri \cite{neiheiser2021kauri},
Themis \cite{kelkar2021themis}, Tendermint \cite{buchman2018latest}, Prime \cite{amir2011prime},
PoE \cite{gupta2021proof}, CheapBFT \cite{kapitza2012cheapbft},
Q/U \cite{abd2005fault}, FaB \cite{martin2006fast}, and Zyzzyva \cite{kotla2007zyzzyva}.

\item {\bf Evolve an existing protocol to derive new variants.}
Second, a key benefit of \sys is in evolving existing protocols to propose new variants.
This incremental design paradigm allows one to adapt a BFT implementation over time to suit the deployment environment.
For example, we can derive new protocol variants from the well-known PBFT \cite{castro1999practical} protocol using a subset of design choices. 

\item {\bf Uncover new protocols.} Finally, \sys can also enable us to discover
{\em new} protocols in the design space simply by combining different design choices not previously explored.
As a proof of concept, we present two new protocol instances:
a Fast Linear BFT protocol (\flb), that establishes consensus in two linear communication phases, and
a Fast Tree-based balanced BFT protocol (\ftb) that supports load balancing.

\end{itemize}

The paper makes the following contributions.

\begin{itemize}

\item {\bf Unified design space.} A design space for BFT protocols
and a set of design choices is proposed.
The proposed design space captures fundamentally new insights into the characteristics, strengths and weaknesses
of BFT protocols and their design trade-offs.
Moreover, studying the design space of BFT protocols
leads to identifying several plausible points that have not yet been explored.

\item {\bf Unified platform.} We present the design and implementation of {\em \sys},
a tool that aims to unify all BFT protocols within a single platform.
\sys derives valid BFT protocols by combining different design choices.

\item {\bf Implementation and evaluation.}
Within \sys, a wide range of BFT protocols are implemented and evaluated.
This unified deployment and experimentation environment provides new opportunities to evaluate and compare different existing BFT protocols in a fair and more efficient manner
(e.g., identical programming language, used libraries, cryptographic tools, etc).
\end{itemize}

The rest of this paper is organized as follows.
Section~\ref{sec:framework} introduces \sys.
The design space of \sys and its design choices are presented in Sections~\ref{sec:space} and \ref{sec:design}.
Section~\ref{sec:mapping} maps some of the known BFT protocols and two new BFT protocols, \flb and \ftb,
to the design space.
The implementation of \sys is introduced in Section~\ref{sec:implementation}.
Section~\ref{sec:eval} shows the experimental results,
Section~\ref{sec:related} discusses the related work, and
Section~\ref{sec:conc} concludes the paper.
\section{The \sys Overview}\label{sec:framework}

\noindent {\bf System model.}
A BFT protocol runs on a network consisting of
a set of nodes that may exhibit arbitrary, potentially malicious, behavior.
BFT protocols use the State Machine Replication (SMR) algorithm \cite{lamport1978time,schneider1990implementing}
where the system provides a replicated service whose state is mirrored across different deterministic replicas.
At a high level, the goal of a BFT SMR protocol is to assign each client
request an order in the global service history and execute it in that order \cite{singh2008bft}.
In a BFT SMR protocol,
all non-faulty replicas execute the same requests in the same order ({\em safety}) and
all correct client requests are eventually executed ({\em liveness}).
In an asynchronous system, where replicas can fail,
there are no consensus solutions that guarantees both safety and liveness (FLP result) \cite{fischer1985impossibility}.
As a result, asynchronous consensus protocols rely on different techniques such as
randomization \cite{cachin2005random,ben1983another,rabin1983randomized},
failure detectors \cite{chandra1996unreliable,malkhi1997unreliable},
hybridization/wormholes \cite{correia2005low,neves2005solving}
and partial synchrony \cite{dwork1988consensus, dolev1987minimal}
to circumvent the FLP impossibility result.

\begin{figure}[t]
\centering
\includegraphics[width= 0.3\linewidth]{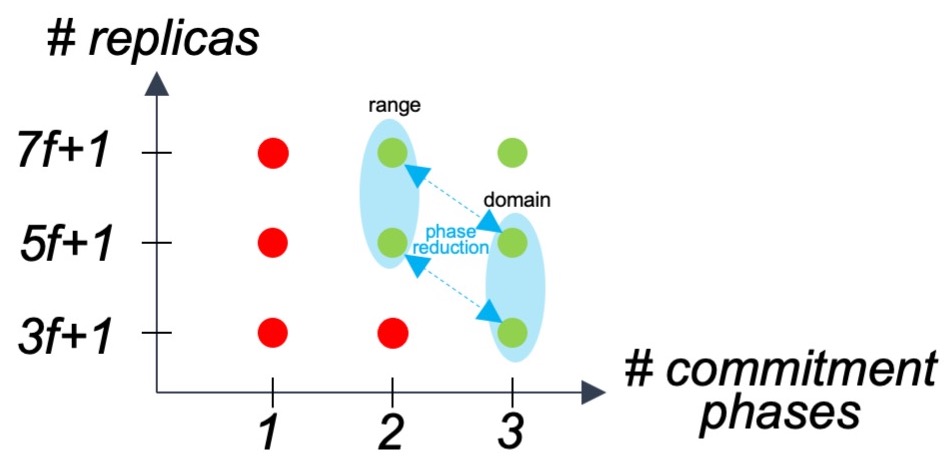}
\caption{A simplified design space with two dimensions:
{\small \sf number of replicas} and {\small \sf number of commitment phases}.
Green dots (\validp{3pt}) specify valid points (i.e., BFT protocols)
while red dots (\invalidp{3pt}) show invalid points (i.e., impossible protocols).
A design choice, i.e., {\small \sf phase reduction (through redundancy)}, is a one-to-one transformation function
that maps a protocol in its domain to another protocol in its range.
}
\label{fig:space}
\end{figure}

\sys assumes the partially synchrony model as it is used in most practical BFT protocols.
In the partially synchrony model, 
there exists an unknown global stabilization time (GST), after which
all messages between correct replicas are received within some unknown bound $\Delta$.
\sys further inherits the standard assumptions of existing BFT protocols.
First, while there is no upper bound on the number of faulty clients,
the maximum number of concurrent malicious replicas is assumed to be $f$.
Second, replicas are connected via an unreliable network that might drop, corrupt, or delay messages.
Third, the network uses point-to-point bi-directional communication channels to connect replicas.
Fourth, the failure of replicas is independent of each other,
where a single fault does not lead to the failure of multiple replicas.
This can be achieve by either diversifying replica implementation (e.g., n-version programming) \cite{avizienis1985n,forrest1997building} or
placing replicas at different geographic locations (e.g., datacenters) \cite{berger2019resilient,sousa2015separating,veronese2010ebawa,eischer2018latency}.
Finally, a strong adversary can coordinate malicious replicas and
delay communication.
However, the adversary cannot subvert cryptographic assumptions.

\noindent {\bf Usage model.}
\sys aims to help application developers experimentally analyze BFT protocols within one unified platform and
find the BFT protocol that fits the characteristics of their applications.
To achieve this goal, the \sys tool makes available the design dimensions of BFT protocols and different design choices,
i.e., trade-offs between dimensions, to application developers to tune.
Figure~\ref{fig:space} illustrates an example
highlighting the relation between design space, dimensions, design choices, and protocols in \sys.
For the sake of simplicity, we present only two dimensions of the design space, i.e.,
{\small \sf number of replicas} and {\small \sf number of commitment phases}
(the design space of \sys consists of more than $10$ dimensions as described in Section~\ref{sec:space}).
Each dimension, e.g., {\small \sf number of replicas}, can take different values, e.g., $3f+1$, $5f+1$, $7f+1$, etc.
A BFT protocol is then a point in this design space, e.g., $(3, 3f+1)$.
Note that each dimension not presented in this figure also takes a value, e.g.,
{\small \sf communication strategy} is assumed to be {\small \sf pessimistic}.

Moreover, a subset of points is valid and represents BFT protocols.
In Figure~\ref{fig:space}, green dots (\validp{3pt}) specify valid points (i.e., BFT protocols)
while red dots (\invalidp{3pt}) show invalid points (i.e., impossible protocols).
For example, there is no (pessimistic) BFT protocol with $3f+1$ nodes that commits requests in a single commitment phase.
A design choice (Section~\ref{sec:design}) is then a one-to-one function
that maps a BFT protocol in its domain to another protocol in its range.
For example, {\small \sf phase reduction (through redundancy)}
maps a BFT protocol with $3f+1$ nodes and $3$ phases of communication, e.g., PBFT \cite{castro1999practical},
to a BFT protocol with $5f+1$ nodes and $2$ phases of communication, e.g., FaB \cite{martin2006fast}
(assuming both protocols are pessimistic and follow clique topology).
The domain and range of each design choice is a subset of BFT protocols in the design space.

\begin{figure}[h]
\centering
\includegraphics[width= 0.5\linewidth]{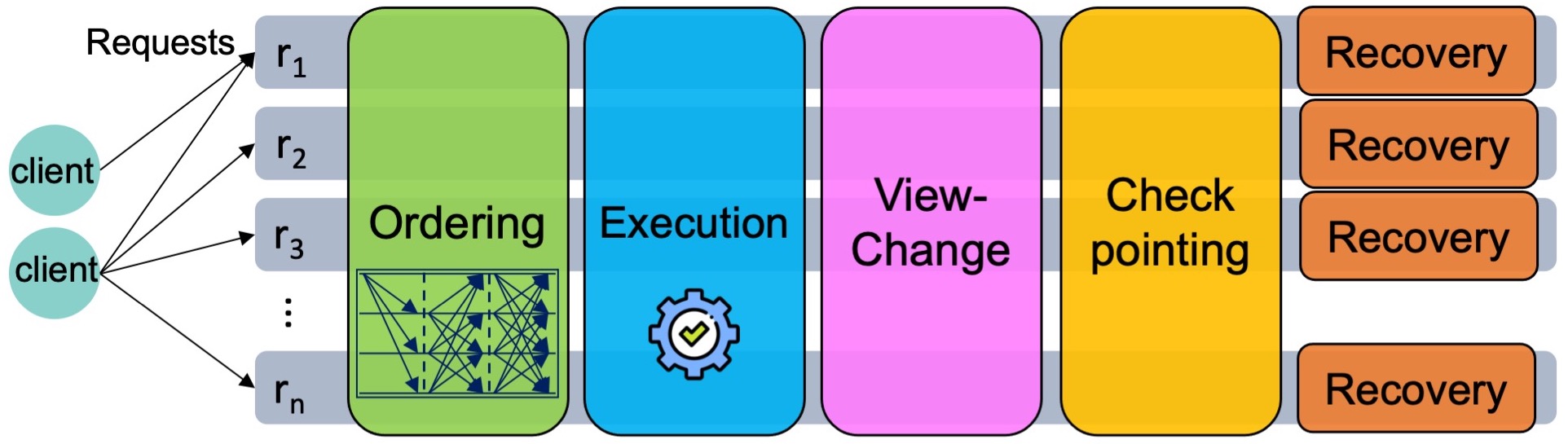}
\caption{Different stages of replicas in a BFT protocol}
\label{fig:stages}
\end{figure}

In a BFT protocol, as presented in Figure~\ref{fig:stages},
clients communicate with a set of replicas that maintain a copy of the application state.
A replica's lifecycle consists of ordering, execution, view-change, checkpointing, and recovery stages.
The goal of the ordering stage is to establish agreement on a unique order,
among requests executing on the application state. 
In leader-based consensus protocols, which are the focus of this paper, a designated {\em leader} node proposes the order, and to ensure fault tolerance, needs to get agreement from a subset of the nodes, referred to as a {\em quorum}.
In the execution stage, requests are applied to the replicated state machine.
The view-change stage replaces the current leader.
Checkpointing is used to garbage-collect data and enable trailing replicas to catch up, and finally,
the recovery stage recovers replicas from faults by applying software rejuvenation techniques.

\section{Design Space}\label{sec:space}

In \sys, each BFT protocol can be analyzed along several dimensions.
These dimensions (and values associated with each dimension) collectively help
to define the overall design space of BFT protocols supported by \sys.
The dimensions are categorized into four families:
{\em environmental settings} and {\em protocol structure} that present the core dimensions of BFT protocols
and are shared among all BFT protocols, a set of optional {\em QoS features}
including order-fairness and load balancing that a BFT protocol might support, and
a set of {\em performance optimizations} for tuning BFT protocols.
In the rest of this section, we describe these families of dimensions in greater detail. As we describe each dimension, we prefix label them with "E" for environmental settings, "P" for protocol structure, etc. Hence, "E 1" refers to the first dimension in the environmental settings dimensions family.

This section is not meant to provide a fully exhaustive set of dimensions, but rather to demonstrate the overall methodology used to define dimensions usable in \sys.

\subsection{Environmental Settings}

Environmental settings broadly speaking encompass
the deployment environment for a BFT protocol, e.g., network size.
These input parameters help scope the class of BFT protocols that can be supported to best fit each deployment environment.

\begin{infra}
\label{dim:nodes}
{\bf Number of replicas.} Our first dimension concerns selecting BFT protocols based on the number of replicas
(i.e., network and quorum size) used in a deployment.
In the presence of $f$ malicious failures, BFT protocols require at least $3f{+}1$ replicas to
guarantee safety \cite{lamport1982byzantine, bracha1984asynchronous, bracha1985asynchronous, dwork1988consensus,correia2006consensus}.
Using trusted hardware, however, the malicious behavior of replicas can be restricted.
Hence, $2f+1$ replicas are sufficient to guarantee safety
\cite{chun2007attested, veronese2013efficient, veronese2010ebawa, reiser2007hypervisor, correia2004tolerate,veronese2010ebawa,correia2013bft}.
Prior proposals on reducing the required number of replicas to $2f+1$ \cite{aguilera2018passing,aguilera2019impact,aguilera2021frugal} involve either leveraging new hardware capabilities or using message-and-memory models. Increasing the number of replicas to $5f+1$ \cite{martin2006fast} (its proven lower bound, $5f-1$ \cite{kuznetsov2021revisiting,abraham2021good})
or $7f+1$ \cite{song2008bosco}, on the other hand, reduces the number of communication phases.
A BFT protocol might also optimistically assume the existence of a quorum of $2f+1$ active non-faulty replicas
to establish consensus \cite{kapitza2012cheapbft,distler2016resource}.
Using both trusted hardware and active/passive replication, the quorum size is further reduced to $f+1$
during failure-free situations \cite{distler2016resource,distler2011spare,kapitza2012cheapbft}.
\end{infra}

\begin{infra}
\label{dim:topology}
{\bf Communication topology.}
\sys allows users to analyze BFT protocols based on communication topologies including:
(1) the star topology where communication is strictly from a designated replica, e.g., the leader,
to all other replicas and vice-versa, resulting in linear message complexity \cite{yin2019hotstuff,kotla2007zyzzyva},
(2) the clique topology where all (or a subset of) replicas communicate directly with each other resulting in quadratic message complexity \cite{castro1999practical},
(3) the tree topology where the replicas are organized in a tree
with the leader placed at the root, and
at each phase, a replica communicates with either its child replicas or its parent replica causing logarithmic message complexity \cite{neiheiser2021kauri,kokoris2019robust,kogias2016enhancing}, or
(4) the chain topology where replicas construct a pipeline and each replica communicates with its successor replica (constant message complexity) \cite{aublin2015next}.
\end{infra}

\begin{infra}
\label{dim:auth}
{\bf Authentication.}
Participants authenticate their messages to enable other replicas to verify a message's origin.
\sys support both signatures, e.g., RSA \cite{rivest2019method}
and authenticators \cite{castro1999practical}, i.e., vectors of message authentication codes (MACs) \cite{tsudik1992message}.
Constant-sized threshold signatures \cite{shoup2000practical,cachin2005random}
have also been used to reduce the size of a set (quorum) of signatures.
Signatures are typically more costly than MACs.
However, in contrast to MACs, signatures provide non-repudiation and
are not vulnerable to MAC-based attacks from malicious clients.
A BFT protocol might even use different techniques (i.e., signatures and MACs) in different stages
to authenticate messages sent by clients, by replicas in the ordering stage, and by replicas during view-change.
\end{infra}

\begin{infra}
\label{dim:timers}
{\bf Responsiveness, Synchronization and Timers.}
A BFT protocol is {\em responsive} 
if its normal case commit latency depends only on the actual network delay needed for replicas to process and exchange messages
rather than any (usually much larger) predefined upper bound
on message transmission delay \cite{pass2017hybrid, pass2018thunderella, attiya1994bounds, shrestha2020optimality}.
Responsiveness might be sacrificed in different ways.
First, when the rotating leader mechanism is used,
the new leader might need to wait for a predefined time 
before initiating the next request to ensure that
it receives the decided value from all non-faulty but slow replicas,
e.g., Tendermint \cite{kwon2014tendermint} and Casper \cite{buterin2017casper}.
Second, optimistically assuming all replicas are non-faulty,
replicas (or clients) need to wait for a predefined upper bound to receive messages from all replicas,
e.g., SBFT \cite{gueta2019sbft} 
and Zyzzyva \cite{kotla2007zyzzyva}. 
\end{infra}

BFT protocols need to guarantee that all non-faulty replicas will eventually be synchronized
to the same view with a non-faulty leader
enabling the leader to collect the decided values in previous views and
making progress in the new view \cite{naor2020expected,naor2019cogsworth,bravo2020making}.
This is needed because a quorum of $2f+1$ replicas might include $f$ Byzantine replicas and
the remaining $f$ "slow" non-faulty replicas might stay behind (i.e., {\em in-dark}) and may not even advance views at all.
{\em View synchronization} can be achieved in different ways such as
integrating the functionality with the core consensus protocol, e.g., PBFT \cite{castro1999practical}, or
assigning a distinct synchronizer component, e.g., Pacemaker in HotStuff \cite{yin2019hotstuff}, and
hardware clocks \cite{abraham2019synchronous}.

Depending on the environment, network characteristics, and processing strategy, BFT protocols use different timers to ensure responsiveness and synchronization.
Protocols can be configured with the following timers by \sys to achieve these goals.
\begin{enumerate}
    \item[$\tau_1$.] Waiting for reply messages, e.g., Zyzzyva \cite{kotla2007zyzzyva}, 
    \item[$\tau_2$.] Triggering (consecutive) view-change, e.g., PBFT \cite{castro1999practical},
    \item[$\tau_3$.] Detecting backup failures, e.g., SBFT \cite{gueta2019sbft},
    \item[$\tau_4$.] Quorum construction in an ordering phase,
    e.g., {\sf \small prevote} and {\sf \small precommit} timeouts in Tendermint \cite{buchman2016tendermint},
    \item[$\tau_5$.] Synchronization for view change, e.g., Tendermint \cite{buchman2016tendermint},
    \item[$\tau_6$.] Finishing  a (preordering) round, e.g., Themis \cite{kelkar2021themis},
    \item[$\tau_7$.] Performance check (heartbeat timer), e.g., Aardvark \cite{clement2009making},
    \item[$\tau_8$.] Atomic recovery (watchdog timer) to periodically hand control to a recovery monitor \cite{castro2000proactive}, e.g., PBFT \cite{castro2002practical}.
\end{enumerate}

\subsection{Protocol Structure}

Our next family of dimensions concerns customization of the protocol structure by \sys,
which will further define the class of protocols permitted.

\begin{struct}
\label{dim:strategy}
{\bf Commitment strategy.}
\sys supports BFT protocols that process transactions in either an optimistic, pessimistic, or robust manner.
{\em Optimistic} BFT protocols make optimistic assumptions on failures, synchrony, or data contention
and might execute requests without necessarily establishing consensus.
An optimistic BFT protocol might make a subset of the following assumptions:
\begin{enumerate}
\item[$a_1$.] \ul{The leader is non-faulty},
assigns a correct order to requests and sends it to all backups, e.g., Zyzzyva \cite{kotla2007zyzzyva},
\item[$a_2$.] \ul{The backups are non-faulty}
and {\em actively} and {\em honestly} participate in the protocol, e.g., CheapBFT\cite{kapitza2012cheapbft},
\item[$a_3$.] \ul{All non-leaf replicas in a tree topology are non-faulty}, e.g., Kauri\cite{neiheiser2021kauri},
\item[$a_4$.] \ul{The workload is conflict-free}
and concurrent requests update disjoint sets of data objects, e.g., Q/U \cite{abd2005fault},
\item[$a_5$.] \ul{The clients are honest}, e.g., Quorum \cite{aublin2015next}, and
\item[$a_6$.] \ul{The network is synchronous} (in a time window),
and messages are not lost or highly delayed, e.g., Tendermint \cite{buchman2016tendermint}. 
\end{enumerate}

Optimistic protocols are classified into speculative and non-speculative protocols.
In non-speculative protocols, e,g., SBFT \cite{gueta2019sbft} and CheapBFT\cite{kapitza2012cheapbft},
replicas execute a transaction only if the optimistic assumption holds.
Speculative protocols, e.g., Zyzzyva \cite{kotla2007zyzzyva} and PoE \cite{gupta2021proof}, on the other hand,
optimistically execute transactions. If the assumption is not fulfilled, replicas might have to
rollback the executed transactions.
Optimistic BFT protocols improve performance in fault-free situations.
If the assumption does not hold, the replicas, e.g., SBFT \cite{gueta2019sbft},
or clients, e.g., Zyzzyva \cite{kotla2007zyzzyva}, detect the failure and use the fallback protocol.

{\em Pessimistic} BFT protocols, on the other hand,
tolerate the maximum number of possible concurrent failures $f$ without making any
assumptions on failures, synchrony, or data contention.
In pessimistic BFT protocols, replicas communicate to agree on
the order of requests.
Finally, {\em robust} protocols, e.g.,
Prime \cite{amir2011prime}, Aardvark \cite{clement2009making}, R-Aliph \cite{aublin2015next},
Spinning \cite{veronese2009spin} and RBFT \cite{aublin2013rbft},
go one step further and consider scenarios where the system is under attack.

In summary, BFT protocols demonstrate different performance in failure-free, low failure, and under attack situations.
Optimistic protocols deliver superior performance in failure-free situations.
However, in the presence of failure, their performance is significantly reduced especially
when the system is under attack.
On the other hand, pessimistic protocols provide high performance in failure-free situations and
are able to handle low failures with acceptable overhead.
However, they show poor performance when the system is under attack.
Finally, robust protocols are designed for under-attack situations and demonstrate moderate performance in all three situations.
\end{struct}

\begin{struct}
\label{dim:phases}
{\bf Number of commitment phases.}
The number of commitment (ordering) phases or {\em good-case latency} \cite{abraham2021good} of a BFT SMR protocol
is the number of phases needed for all non-faulty replicas to commit when the leader is non-faulty, and the network is synchronous.
We consider the number of commitment phases from the first time a replica (typically the leader)
receives a request to the first time any participant
(i.e., leader, backups, client) learns the commitment of the request,
e.g., PBFT executes in 3 phases.
\end{struct}

\begin{struct}
\label{dim:view}
{\bf View-change.}
BFT protocols follow either the {\em stable leader}  or
the {\em rotating leader} mechanism to replace the current leader.
The stable leader mechanism \cite{castro1999practical,gueta2019sbft,martin2006fast,kotla2007zyzzyva}
replaces the leader when the leader is suspected to be faulty by other replicas.
In the rotating leader mechanism \cite{aiyer2005bar,kwon2014tendermint,yin2019hotstuff,chan2020streamlet,chan2018pala,chan2018pili,gilad2017algorand, hanke2018dfinity,kokoris2019robust,veronese2010ebawa,veronese2009spin,clement2009making,buchnik13fireledger},
the leader is replaced periodically,
e.g., after a single attempt, insufficient performance,
or an epoch (multiple requests).

Using the stable leader mechanism, the view-change stage becomes more complex.
However, the routine is only executed when the leader is suspected to be faulty.
On the other hand, the rotating leader mechanism requires
ensuring view synchronization frequently (whenever the leader is rotated).
Rotating the leader has several benefits such as
balancing load across replicas \cite{veronese2009spin, behl2015consensus,behl2017hybrids},
improving resilience against slow replicas \cite{clement2009making}, and
minimizing communication delays between clients and the leader \cite{veronese2010ebawa,mao2009towards,eischer2018latency}.
\end{struct}

\begin{struct}
\label{dim:checkpoint}
{\bf Checkpointing.}
The checkpointing mechanism is used to
first, garbage-collect data of completed consensus instances to save space and
second, restore in-dark replicas (due to network unreliability or leader maliciousness)
to ensure all non-faulty replicas are up-to-date \cite{gupta2021proof,distler2021byzantine,castro1999practical}.
The checkpointing stage typically is initiated after a fixed checkpoint window
in a decentralized manner without relying on a leader \cite{castro1999practical}.
\end{struct}

\begin{struct}
\label{dim:recovery}
{\bf Recovery.}
When there are more than $f$ failures, BFT protocols, apart from some exceptions \cite{li2007beyond,chun2007attested},
completely fail and do not give any guarantees on their behavior \cite{distler2021byzantine}.
BFT protocols perform recovery using {\em reactive} or {\em proactive} mechanisms (or a combination~\cite{sousa2009highly}).
Reactive recovery mechanisms detect faulty replica behavior \cite{haeberlen2006case} and
recover the replica by applying software rejuvenation techniques \cite{cotroneo2014survey,huang1995software} where
the replica reboots, reestablishes its connection with other replicas and clients, and updates its state. 
On the other hand, proactive recovery mechanisms recover replicas in periodic time intervals.
Proactive mechanisms do not require any fault detection techniques, however, they
might unnecessarily recover non-faulty replicas \cite{distler2021byzantine}.
During recovery, a replica is unavailable.
A BFT protocol can rely on $3f+ 2k+1$ replicas to improve resilience and availability during recovery
where $k$ is the maximum number of servers that rejuvenate concurrently \cite{sousa2009highly}.
To prevent attackers from disrupting the recovery process,
each replica requires
a trusted component, e.g., secure coprocessor \cite{castro2002practical},
a synchronous wormhole \cite{verissimo2006travelling} or a virtualization layer \cite{reiser2007hypervisor,distler2011spare},
that remains operational even if the attacker controls the replica and
a read-only memory that an attacker cannot manipulate.
The memory content remains persistent (e.g., on disk) across machine reboots and
includes all information needed for bootstrapping a correct replica after restart \cite{distler2021byzantine}.
\end{struct}

\begin{struct}
\label{dim:client}
{\bf Types of Clients.}
\sys supports three types of clients: requester, proposer, and repairer.
{\em Requester} clients perform a basic functionality and communicate with replicas by sending requests and receiving replies.
A requester client might need to verify the results by waiting for a number of matching replies, e.g.,
$f+1$ in PBFT \cite{castro1999practical}, $2f+1$ in PoE \cite{gupta2021proof} and PBFT \cite{castro1999practical} (for read-only requests) , or $3f+1$ is Zyzzyva \cite{kotla2007zyzzyva}. 
Using trusted components, e.g., Troxy \cite{li2018troxy}, or
threshold signatures, e.g., SBFT \cite{gueta2019sbft},
the client does not even need to wait for and verify multiple results from replicas. 
Clients might also play the {\em proposer} role by
proposing a sequence number (acting as the leader) for its request \cite{malkhi1998byzantine,malkhi1998survivable,guerraoui2010next,abd2005fault}.
{\em Repairer} clients, on the other hand, detect the failure of replicas, e.g., Zyzzyva \cite{kotla2007zyzzyva}, or
even change the protocol configuration,
e.g., Scrooge\cite{serafini2010scrooge}, Abstract \cite{aublin2015next}, and Q/U\cite{abd2005fault}.
\end{struct}

\subsection{Quality of Service}

There are some optional QoS features that \sys can analyze. We list two example dimensions.

\begin{feature}
\label{dim:fairness}
{\bf Order fairness.}
Order-fairness deals with preventing adversarial manipulation of request ordering \cite{zhang2020byzantine,kursawe2020wendy, kursawe2021wendy,kelkar2020order,kelkar2021themis,baird2016swirlds}.
Order-fairness is defined as: "if a large number of replicas receives a
request $t_1$ before another request $t_2$, then $t_1$ should be ordered before $t_2$" \cite{kelkar2020order}.
Order fairness has been partially addressed using different techniques:
(1) monitoring the leader to ensure it does not initiate two new requests from the same client
before initiating an old request of another client, e.g., Aardvark \cite{clement2009making},
(2) adding a preordering phase, e.g., Prime \cite{amir2011prime},
where replicas order the received requests locally and share their own ordering with each other,
(3) encrypting requests and revealing the contents only once their ordering is fixed \cite{asayag2018fair,cachin2001secure,miller2016honey,stathakopoulou2021adding},
(4) reputation-based systems \cite{asayag2018fair,kokoris2018omniledger,lev2020fairledger,crain2021red} to detect unfair censorship of specific client requests, and
(5) providing opportunities for every replica to propose and commit its requests using fair election \cite{kiayias2017ouroboros,abraham2018solida,gilad2017algorand,lev2020fairledger,pass2017hybrid,asayag2018fair,yakira2021helix}.
\end{feature}

\begin{feature}
\label{dim:load}
{\bf Load balancing.}
The performance of fault-tolerant protocols is usually limited 
by the computing and bandwidth capacity of the leader \cite{biely2012s,moraru2013there,ailijiang2019dissecting,charapko2021pigpaxos,neiheiser2021kauri,moraru2012egalitarian,whittaker2021scaling}.
The leader coordinates the consensus protocol and multicasts/collects messages
to all other replicas in different protocol phases.
Load balancing is defined as distributing the load among the replicas of the system to balance
the number of messages any single replica has to process.

Load balancing can be partially achieved using the rotating leader mechanism, multi-layer, or multi-leader BFT protocols.
When the rotating leader mechanism is used, one (leader) replica still is highly loaded in each consensus instance.
In multi-layer BFT protocols \cite{li2020scalable,amir2008steward,neiheiser2018fireplug,gupta2020resilientdb,nawab2019blockplane}
the load of the leader is distributed between the leaders of different clusters.
However, the system still suffers from load imbalance between the leader and backup replicas in each cluster.
In multi-leader protocols \cite{arun2019ezbft,voron2019dispel,stathakopoulou2019mir,gupta2021rcc,alqahtani2021bigbft,avarikioti2020fnf},
all replicas can initiate consensus to partially order requests in parallel.
However, slow replicas still affect the global ordering of requests.
\end{feature}


\subsection{Performance Optimization Dimensions}\label{append:opt}

Finally, we present a set of optimization dimensions that target the performance of a BFT protocol.

\begin{optim}
\label{optim:order}
{\bf Out-of-order processing.}
The out-of-order processing mechanism enables the leader to continuously
propose new requests even when previous requests are still being processed by the backups \cite{gupta2021proof}.
Out-of-order processing of requests is possible if the leader does not need to include any certificate
or hash of the previous request (block) in its next request.
\end{optim}

\begin{optim}
\label{optim:pipe}
{\bf Request pipelining.}
Using request pipelining,
the messages of a new consensus instance are piggybacked on the second
round messages of the previous instance \cite{yin2019hotstuff, neiheiser2021kauri}.
This technique is especially efficient when a protocol rotates the leader after every consensus instance.
\end{optim}

\begin{optim}
\label{optim:para-ord}
{\bf Parallel ordering.}
Client requests can be ordered in parallel by relying on a set of independent ordering groups \cite{behl2015consensus,behl2017hybrids,li2016sarek}
where each group orders a subset of client requests and then all results are deterministically merged into the final order.
Similarly, in multi-leader protocols \cite{arun2019ezbft,voron2019dispel,stathakopoulou2019mir,gupta2021rcc,alqahtani2021bigbft,avarikioti2020fnf,li2016sarek,milosevic2013bounded,eischer2019scalable,voron2019dispel,arun2022scalable}, different replicas are designated as the leader for different consensus instances in parallel and then a global order is determined.
\end{optim}

\begin{optim}
\label{optim:para-exe}
{\bf Parallel execution.}
Transactions can be executed in parallel to improve the system's overall performance.
One approach is to detect non-conflicting transactions and execute them in parallel \cite{escobar2019boosting,kotla2004high,amiri2019parblockchain}.
This approach requires {\em a priori} knowledge of a transaction's read-set and write-set.
Switching the order of agreement and execution stages and optimistically executing transactions in parallel is another approach~\cite{kapritsos2012all,androulaki2018hyperledger}.
If the execution results are inconsistent
(due to faulty replicas, conflicting transactions, or nondeterministic execution),
replicas need to rollback their states and sequentially and deterministically re-execute the requests.
switching the order of agreement and execution stages also enables replicas
to detect any nondeterministic execution \cite{kapritsos2012all,androulaki2018hyperledger}.
\end{optim}

\begin{optim}
\label{optim:readonly}
{\bf Read-only requests processing.}
In pessimistic protocols, replicas can directly execute read-only requests without establishing consensus.
However, since replicas may execute the read requests on different states,
even non-faulty replicas might not return identical results.
To resolve this, the number of required matching replies for both normal and read-only requests
needs to be increased from $f+1$ to $2f+1$ in order to
ensure consistency (i.e., quorum intersection requirement) \cite{castro2002practical}.
This, however, results in a liveness challenge because $f$ non-faulty replicas might be slow (or in-dark) and
not receive the request. As a result, the client might not be able to collect $2f+1$ matching responses
(since Byzantine replicas may not send a correct reply to the client).
\end{optim}

\begin{optim}
\label{optim:separate}
{\bf Separating ordering and execution.}
The ordering and execution stages can be separated and implemented in different processes.
This separation leads to several advantages \cite{distler2021byzantine} such as
preventing malicious execution replicas from leaking confidential application state to clients \cite{yin2003separating,duan2016practical},
enabling large requests to bypass the ordering stage \cite{clement2009upright},
moving application logic to execution virtual machine \cite{distler2011spare,wood2011zz,reiser2007hypervisor} or
simplifying the parallel ordering of requests  \cite{behl2015consensus,ben2019scalable}.
Moreover, while $3f{+}1$ replicas are needed for ordering, $2f+1$ replicas are sufficient to execute transactions \cite{yin2003separating}.
\end{optim}

\begin{optim}
\label{optim:trusted}
{\bf Trusted hardware.}
Using Trusted execution environments (TEEs) such as Intel's SGX \cite{mckeen2013innovative}, Sanctum \cite{costan2016sanctum},
and Keystone \cite{lee2020keystone},
the number of required replicas can be lowered to $2f+1$ because
the trusted component prevents a faulty replica from sending conflicting
messages to different replicas without being detected.
A trusted component may include
an entire virtualization layer \cite{veronese2010ebawa, distler2011spare,reiser2007hypervisor},
a multicast ordering service executed on a hardened Linux kernel \cite{correia2007worm,correia2004tolerate},
a centralized configuration service \cite{renesse2012byzantine},
a trusted log \cite{chun2007attested},
a trusted platform module, e.g., counter \cite{veronese2013efficient,veronese2010ebawa},
a smart card TrInc \cite{levin2009trinc},
or an FPGA \cite{distler2016resource,kapitza2012cheapbft}.
\end{optim}

\begin{optim}
\label{optim:client}
{\bf Request/reply dissemination.}
A client can either multicast its request to {\em all replicas} \cite{bessani2014state,veronese2009spin,cowling2006hq} 
where each replica relays the request to the leader
or
optimistically send its request to a {\em contact replica}, typically the leader.
The contact replica is known to the client through a reply to an earlier request
\cite{castro1999practical,kotla2007zyzzyva}.
If the client timer for the request ($\tau_1$) expires, the client multicasts its request to all replicas.
This optimistic mechanism requires fewer messages to be sent from clients to the replicas. However,
this comes at the cost of increased network traffic between replicas, because the leader needs to
disseminate the full request to other replicas to enable them to eventually execute it.

On the other hand, all replicas can send the results to clients in their reply messages.
This, however, leads to significant network overhead for large results.
A protocol can optimistically rely on a designated responder replica (chosen by the client or servers)
to send the full results.
Other replicas then either send the hash of the results to the client or
send a signed message to the responder enabling the responder to generate a proof for the results, e.g., SBFT \cite{gueta2019sbft}.
While this technique reduces network overhead, the client might not receive the results
if the responder replica is faulty, the network is unreliable,
or the responder replica was in-dark and skipped the execution and applied a checkpoint to catch up \cite{distler2021byzantine}.
\end{optim}


\section{Design Choices Landscape}\label{sec:design}

Given a set of specified dimension values in Section~\ref{sec:space},
\sys generates a set of valid protocols that meet a user query.
Each protocol represents a point in the \sys design space.
In this section, using the classical PBFT~\cite{castro1999practical,castro2002practical}
as a driving example, we demonstrate how different points in the design space lead to different trade-offs.
Each design choice is a {\em one-to-one function}
that maps a valid input point (i.e., a BFT protocol) to another valid output point in the design space.
The domain of each function (design choice) is a subset of the valid points in the design space.
These design choices enable \sys to generate valid BFT protocols.

\subsection{Background on PBFT}

\begin{figure}[t]
\centering
\includegraphics[width= 0.5\linewidth]{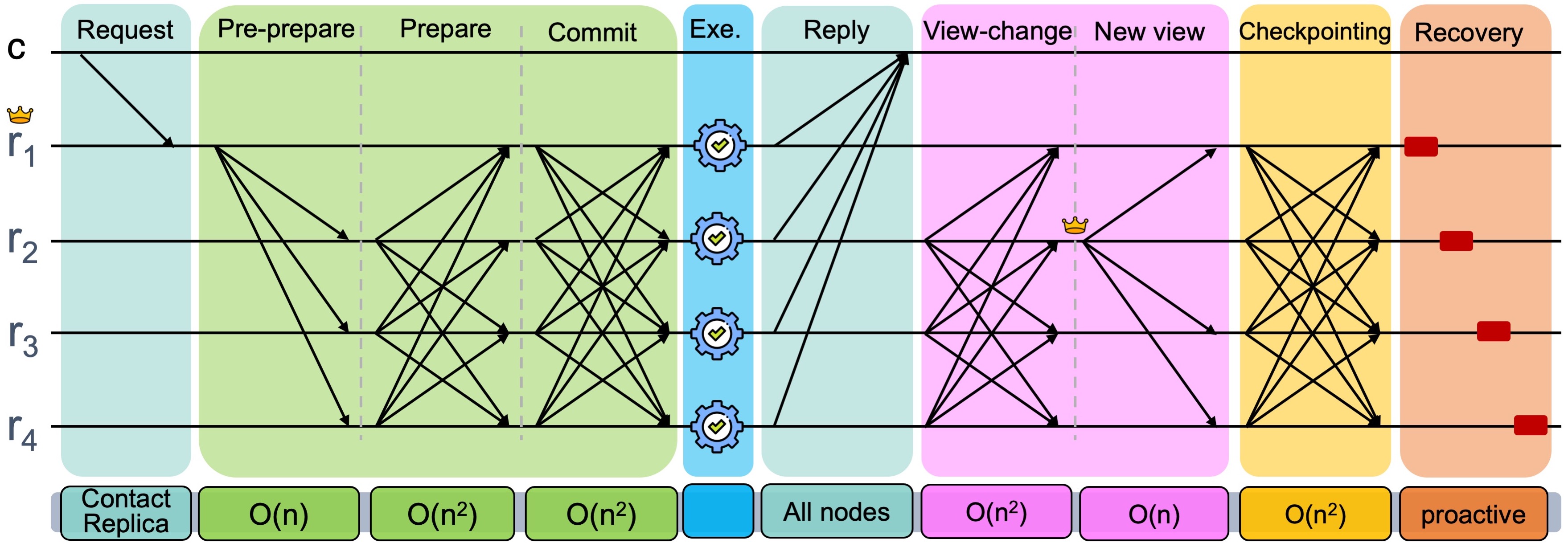}
\caption{Different stages of PBFT protocol}
\label{fig:pbft}
\end{figure}

PBFT, as shown in Figure~\ref{fig:pbft}, is a leader-based protocol that operates in a 
succession of configurations called {\em views} \cite{el1985efficient,el1985availability}.
Each view is coordinated by a {\em stable} leader (primary) and
the protocol {\em pessimistically} processes requests.
In PBFT, the number of replicas, $n$, is assumed to be $3f+1$ and the ordering stage 
consists of \one, \two, and \three phases.
The \one phase assigns an order to the request,
the \two phase guarantees the uniqueness of the assigned order and
the \three phase guarantees that the next leader can assign order in a safe manner.

During a normal case execution of PBFT,
clients send their signed \req messages to the leader.
In the \one phase, the leader 
assigns a sequence number to the request to determine the execution order of the request
and multicasts a \one message including the {\em full request} to all {\em backups}.
Upon receiving a valid \one message from the leader,
each backup replica multicasts a \two message to all replicas
and waits for \two messages from $2f$ different replicas (including the replica itself) that match the \one message.
The goal of the {\sf \small pre-prepare} phase is to guarantee safety within the view, i.e.,
a majority of non-faulty replicas received matching \one messages from the leader replica and agree with the order of the request.
Each replica then multicasts a \three message to all replicas.
Once a replica receives $2f+1$ valid \three messages from different replicas including itself)
that match the \one message, it commits the transaction.
The goal of the \three phase is to ensure safety across views, i.e., 
the request has been replicated on a majority of non-faulty replicas.
The second and third phases of PBFT follow the {\em clique} topology, i.e., have $\mathcal{O}(n^2)$ message complexity.
If the replica has executed all requests with lower sequence numbers, it executes the transaction and
sends a \reply to the client.
The client waits (timer $\tau_1$) for $f+1$ matching results from different replicas.

In the view change stage, upon detecting the failure of the leader of view $v$ using timeouts (timer $\tau_2$),
backups exchange {\sf \small view-change} messages including
transactions that have been received by the replicas.
After receiving $2f+1$ {\sf \small view-change} messages,
the designated stable leader of view $v+1$ (the replica with \texttt{ID} = $v+1$ \texttt{mod} $n$)
proposes a {\sf \small new view}
message including a list of transactions that should be processed in the new view.

In PBFT replicas periodically generate and send {\sf \small checkpoint} messages to all other replicas.
If a replica receives $2f+1$ matching  {\sf \small checkpoint} messages, the checkpoint is stable.
PBFT also includes a {\em proactive} recovery mechanism that periodically (timer $\tau_8$) rejuvenates replicas one by one.

PBFT uses either signatures \cite{castro1999practical} or MACs \cite{castro2002practical} for authentication.
Using MACs, replicas need to send {\sf \small view-change-ack} messages to the leader after receiving {\sf \small view-change} messages.
Since {\sf \small new view} messages are not signed, these {\sf \small view-change-ack} enable replicas
to verify the authenticity of {\sf \small new view} messages.

\subsection{Expanding the Design Choices of PBFT}\label{sec:choices}

Using the PBFT protocol and our design dimensions as a baseline,
we illustrate a series of design choices
that expose different trade-offs BFT protocols need to make.
Each design choice acts as 
a one-to-one function that changes the value of one or multiple dimensions to
map a BFT protocol, e.g., PBFT, to another BFT protocol.

\begin{tr}
\label{tr:phase}
\textbf{Linearization.}
This function explores a trade-off between communication topology and communication phases.
The function takes a quadratic communication phase,
e.g., {\sf \small prepare} or {\sf \small commit} in PBFT,
and split it into two linear phases:
one phase from all replicas to a collector (typically the leader)
and one phase from the collector to all replicas, e.g., SBFT \cite{gueta2019sbft}, HotStuff \cite{yin2019hotstuff}.
The output protocol requires (threshold) signatures for authentication.
The collector collects a quorum of (typically $n-f$) {\em signatures} from other replicas and 
broadcasts its message including the signatures as the certificate of receiving messages to every replica.
Using threshold signatures \cite{shoup2000practical,ramasamy2005parsimonious,cachin2001secure,cachin2005random}
the collector message size can be further reduced from linear to constant.
\end{tr}

\begin{tr}
\label{tr:nodes}
\textbf{Phase reduction through redundancy.}
This function explores a trade-off between the number of ordering phases and the number of replicas.
The function transforms a protocol with $3f+1$ replicas and $3$ ordering phases (i.e., one linear, two quadratic), e.g., PBFT, to
a fast protocol with $5f+1$ replicas and $2$ ordering phases (one linear, one quadratic), e.g., FaB \cite{martin2006fast}.
In the second phase of the protocol, matching messages from a quorum of $4f+1$ replicas are required.
Recently, $5f-1$ has been proven as the lower bound for two-step (fast) Byzantine consensus \cite{kuznetsov2021revisiting,abraham2021good}.
The intuition behind the $5f-1$ lower bound is that in an authenticated model, when replicas detect leader equivocation and
initiate view-change, they do not include view-change messages coming from the malicious leader reducing the maximum number of faulty messages to $f-1$ \cite{kuznetsov2021revisiting,abraham2021good}.
\end{tr}

\begin{tr}
\label{tr:view1}
\textbf{Leader rotation.}
This function replaces the stable leader mechanism with the rotating leader mechanism, e.g., HotStuff \cite{yin2019hotstuff}
where the rotation happens after each request or epoch or due to low performance.
The function eliminates the view-change stage and adds a new quadratic phase or $2$ linear phases
(using the linearization function)
to the ordering stage to ensure that the new leader is aware of the correct state of the system.
\end{tr}

\begin{tr}
\label{tr:view2}
\textbf{Non-responsive leader rotation.}
This function replaces the stable leader mechanism with the rotating leader mechanism
{\em without} adding a new ordering phase (in contrast to design choice~\ref{tr:view1})
while sacrificing responsiveness.
The new leader optimistically assumes that the network is synchronous and
waits for a predefined known upper bound $\Delta$ (Timer $\tau_4$) before initiating the next request.
This is needed to ensure that the new leader is aware of the highest assigned order
to the requests, e.g., Tendermint \cite{kwon2014tendermint,buchman2018latest} and Casper \cite{buterin2017casper}.
\end{tr}

\begin{tr}
\label{tr:active}
\textbf{Optimistic replica reduction.}
This function reduces the number of involved replicas in consensus from $3f+1$ to $2f+1$
while optimistically assuming all $2f+1$ replicas are non-faulty (assumption $a_2$).
In each phase of a BFT protocol, matching messages from a quorum of $2f+1$ replicas is needed.
If a quorum of $2f+1$ non-faulty replicas is identified, they can order (and execute)
requests without the participation of the remaining $f$ replicas.
Those $f$ replicas remain passive and are needed if any of the $2f+1$ active replicas become faulty \cite{kapitza2012cheapbft,distler2016resource}.
Note that $n$ is still $3f+1$.
\end{tr}

\begin{tr}
\label{tr:1phase}
\textbf{Optimistic phase reduction.}
Given a {\em linear} BFT protocol, this function optimistically eliminates two linear phases
(i.e., equal to the quadratic phase {\sf \small prepare})
assuming all replicas are non-faulty, e.g., SBFT\cite{gueta2019sbft}.
The leader (collector) waits for signed messages from all $3f+1$ replicas in the second phase of ordering,
combines signatures and sends a signed message to all replicas.
Upon receiving the signed message from the leader, each replica ensures that all non-faulty replicas
has received the request and agreed with the order.
As a result, the third phase of communication can be omitted and replicas can directly commit the request.
If the leader has not received $3f+1$ messages after a predefined time (timer $\tau_3$),
the protocol fall backs to its slow path and runs the third phase of ordering.
\end{tr}

\begin{tr}
\label{tr:s1phase}
\textbf{Speculative phase reduction.}
This function, similar to the previous one, optimistically eliminates two linear phases
of the ordering stage assuming that non-faulty replicas construct the quorum of responses, e.g., PoE\cite{gupta2021proof}.
Th main difference is that the leader waits for signed messages from only $2f+1$ replicas in the second phase of ordering
and sends a signed message to all replicas.
Upon receiving a message signed by $2f+1$ replicas from the leader, each replica speculatively executes the transaction,
optimistically assuming that either
(1) all $2f+1$ signatures are from non-faulty replicas or
(2) at least $f+1$ non-faulty replicas received the signed message from the leader.
If (1) does not hold, other replicas receive and execute transaction during the view-change. However,
if (2) does not hold, the replica might have to rollback the executed transaction.
\end{tr}

\begin{tr}
\label{tr:2phase}
\textbf{Speculative execution.}
This function eliminates the {\sf \small prepare} and {\sf \small commit} phases
while optimistically assuming that all replicas are non-faulty
(optimistic assumptions $a_1$ and $a_2$), e.g., Zyzzyva \cite{kotla2007zyzzyva}.
Replicas speculatively execute transactions upon receiving them from the leader.
If the client does not receive $3f+1$ matching replies after a predefined time (timer $\tau_1$)
or it receives conflicting messages,
the client detects failures (repairer)
and communicates with replicas to receive $2f+1$ {\sf \small commit} messages
(two linear phases).
\end{tr}

\begin{table*}[t]
\centering
\scriptsize
\caption{Comparing selected BFT protocols based on different dimensions of \sys design space}
\hspace*{-0.7cm}\begin{tabular}{@{}p{1.7cm}|c|c|c|c|c|c|c|C{0.75cm}|c|C{0.35cm}|C{0.45cm}|c@{}}
\toprule
Protocol & 
\begin{tabular}[c]{@{}c@{}} E\ref{dim:nodes}. \\Replicas\end{tabular}  & 
\begin{tabular}[c]{@{}c@{}} E\ref{dim:topology}. \\ Topo.\end{tabular} &
\begin{tabular}[c]{@{}c@{}} E\ref{dim:auth}. \\ Auth. \end{tabular}&
\begin{tabular}[c]{@{}c@{}} E\ref{dim:timers}. \\ Timers \end{tabular}&
\begin{tabular}[c]{@{}c@{}} P\ref{dim:strategy}. \\Strategy\end{tabular} &
\begin{tabular}[c]{@{}c@{}} P\ref{dim:phases}. \\Phases\end{tabular} & 
\begin{tabular}[c]{@{}c@{}} P\ref{dim:view}. \\ V-change \end{tabular}&
\begin{tabular}[c]{@{}c@{}} P\ref{dim:recovery}. \\ Recov. \end{tabular}&
\begin{tabular}[c]{@{}c@{}} P\ref{dim:client}. \\ Client \end{tabular}&
\begin{tabular}[c]{@{}c@{}} Q\ref{dim:fairness}. \\ Fair. \end{tabular}&
\begin{tabular}[c]{@{}c@{}} Q\ref{dim:load}. \\ Load. \end{tabular}&
\begin{tabular}[c]{@{}c@{}}Design\\ Choices\end{tabular} \\ \midrule

PBFT \cite{castro1999practical}  & $3f+1$ & clique & MAC || Sign & $\tau_1$, $\tau_2$, $\tau_8$ & pessimistic&  $3$    & stable & pro. & Req. &
$\square$  & $\square$ &   (\ref{tr:auth}) \\ \midrule

Zyzzyva \cite{kotla2007zyzzyva}   & $3f+1$ & star  & MAC || Sign  & $\tau_1$, $\tau_2$ & optimis: $a_1$, $a_2$|Spec & $1$ $(3)$  & stable &-  & Rep. &
$\square$  & $\square$ &    \ref{tr:2phase}, (\ref{tr:auth})   \\ \midrule

Zyzzyva5 \cite{kotla2007zyzzyva}   & $5f+1$ & star  & MAC || Sign  & $\tau_1$, $\tau_2$ & optimis: $a_1$|Spec & $1$ $(3)$    & stable & - & Rep. &
$\square$  & $\square$ &  \ref{tr:2phase}, \ref{tr:resilience}, (\ref{tr:auth})   \\ \midrule

PoE \cite{gupta2021proof}       & $3f+1$ & star & MAC || T-Sign & $\tau_1$, $\tau_2$ & optimis: $a_2$ |Spec & 3      & stable & - & Req. &
$\square$  & $\square$ &   \ref{tr:phase}, \ref{tr:s1phase}, \ref{tr:auth}  \\ \midrule

SBFT \cite{gueta2019sbft}     & $3f+1$ & star & T-Sign & $\tau_1$, $\tau_2$, $\tau_3$ & optimis: $a_2$& 3 $(5)$     & stable & - & Req. &
$\square$  & $\square$ &    \ref{tr:phase}, \ref{tr:1phase}, \ref{tr:auth} \\ \midrule

HotStuff \cite{yin2019hotstuff}   & $3f+1$ & star  & T-Sign & $\tau_1$, $\tau_2$ &pessimistic& $7$      & rotating & - & Req. &
$\square$  & $\square$ &    \ref{tr:phase}, \ref{tr:view1}, \ref{tr:auth}   \\ \midrule

Tendermint \cite{buchman2018latest}   & $3f+1$ & clique & Sign & $\tau_1$, $\tau_2$, $\tau_5$, $\tau_6$ & optimis: $a_6$& $3$    & rotating & - & Req. &
$\square$  & $\square$ &    \ref{tr:view2}, \ref{tr:auth}  \\ \midrule

Themis \cite{kelkar2021themis}  & $4f+1$ & star  & T-Sign & $\tau_1$, $\tau_2$, $\tau_6$& pessimistic & $1+7$      & rotating & - & Req. &
$\blacksquare$  & $\square$ &   \ref{tr:phase}, \ref{tr:view1}, \ref{tr:fair}, \ref{tr:auth}   \\ \midrule

Kauri \cite{neiheiser2021kauri}  & $3f+1$ & tree  & T-Sign & $\tau_1$, $\tau_2$ & optimis: $a_3$& $7h$      & stable* & - & Req.  &
$\square$  & $\blacksquare$ &   (\ref{tr:view1}), \ref{tr:load}, \ref{tr:auth}   \\ \midrule

CheapBFT\cite{kapitza2012cheapbft}   & $2f+1$ & clique & MAC & $\tau_1$, $\tau_2$ & optimis: $a_2$&  3 &  stable & - & Req. &
$\square$  & $\square$ &  \ref{tr:active} \\ \midrule

FaB \cite{martin2006fast}        & $5f+1$ & clique  & (Sign) & $\tau_1$, $\tau_2$ & pessimistic& $2$    & stable & - & Req. &
$\square$  & $\square$ &   \ref{tr:nodes}  \\ \midrule

Prime \cite{amir2011prime}   & $3f+1$ & clique  & Sign & $\tau_1$, $\tau_2$, $\tau_6$, $\tau_7$ & robust& 6    & stable & - & Req. &
$\squarelrblack$  & $\square$ & \ref{tr:auth}, \ref{tr:robust} \\ \midrule

Q/U \cite{abd2005fault}   & $5f+1$ & star  & MAC & $\tau_1$, $\tau_2$ & optimis: $a_4$, $a_5$& $1$ $(3)$    & stable & - & Rep. &
$\square$  & $\square$ &   \ref{tr:conflict}, \ref{tr:resilience} \\ \midrule

\rowcolor{teal!20!white}
FLB      & $5f-1$ & clique  & Sign & $\tau_1$, $\tau_2$ & pessimistic& $2$    & stable & - & Req. &
$\square$  & $\square$ &   \ref{tr:phase}, \ref{tr:nodes},  \ref{tr:auth}  \\ \midrule

\rowcolor{teal!20!white}
FTB         & $5f-1$ & tree  & T-Sign & $\tau_1$, $\tau_2$ & optimis: $a_3$& $3h$      & stable & - & Req.  &
$\square$  & $\blacksquare$ &   \ref{tr:phase}, \ref{tr:nodes}, \ref{tr:load}, \ref{tr:auth}   \\ \bottomrule
\end{tabular}
{\scriptsize Hint: T-Sign is used for threshold signatures.
Speculative optimistic protocols are specified by Spec.
The number of phases in slow-path of optimistic protocols is shown within parenthesis.
While Kauri is implemented on top of HotStuff, it does not use the rotating leader mechanism.
Prime provides partial fairness.}
\label{tbl:protocol}
\vspace{-1em}
\end{table*}


\begin{tr}
\label{tr:conflict}
\textbf{Optimistic conflict-free.}
Assuming that requests of different clients are conflict-free (assumption $a_4$),
there is no need for a total order among all transactions.
The function eliminates all three ordering phases
while optimistically assuming that requests are conflict-free and all replicas are non-faulty.
The client becomes the {\em proposer} and sends its request to all (or a quorum of) replicas
where replicas execute the client requests without any further communication \cite{abd2005fault,cowling2006hq}.
\end{tr}

\begin{tr}
\label{tr:resilience}
\textbf{Resilience.}
This function increases the number of replicas by $2f$ to enable the protocol to tolerate $f$ more failure
with the same safety guarantees.
In particular, optimistic BFT protocols that assume all $3f+1$ replicas are non-faulty (quorum size is also $3f+1$)
tolerate zero failures.
By increasing the number of replicas to $5f+1$ replicas, such BFT protocols can
provide the same safety guarantees with quorums of size $4f+1$ while tolerating $f$ failures,
e.g., Zyzzyva5 \cite{kotla2007zyzzyva}, Q/U \cite{abd2005fault}.
Similarly, a protocol with $5f+1$ can tolerates $f$ more faulty replicas by
increasing the network size to $7f+1$ \cite{song2008bosco}.

The function can also be used to provide high availability during the (proactive) recovery stage by
increasing the number of replicas by $2k$ (the quorum size by $k$) where
$k$ is the maximum number of servers that recover in parallel \cite{sousa2009highly}.
\end{tr}

\begin{tr}
\label{tr:auth}
\textbf{Authentication.}
This function replaces MACs with signatures for a given stage of a protocol.
If a protocol follows the star communication topology where a replica needs to include a quorum of signatures
as a proof of its messages, e.g., HotStuff \cite{yin2019hotstuff},
$k$ signatures can be replaced with a single threshold signature.
Note that in such a situation MACs cannot be used (MACs do not provide non-repudiation).
\end{tr}

\begin{figure}
\centering
\includegraphics[width= 0.6\linewidth]{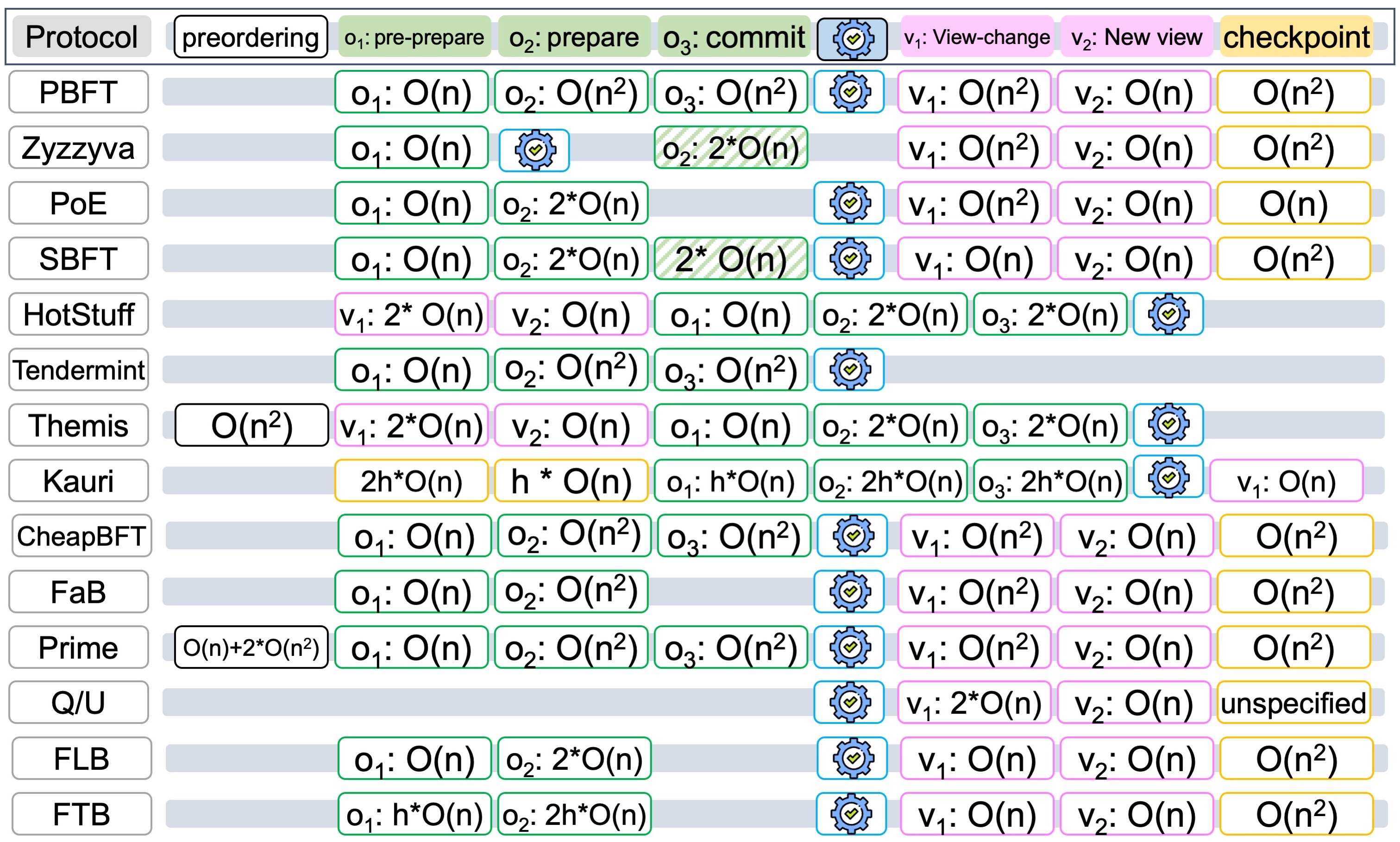}
\caption{Overview of BFT protocols}
\label{fig:protocols}
\vspace{-1em}
\end{figure}

\begin{tr}
\label{tr:robust}
\textbf{Robust.}
This function makes a pessimistic protocol robust
by adding a preordering stage to the protocol, e.g., Prime \cite{amir2011prime}.
In the preordering stage and, upon receiving a request, each replica locally orders and broadcasts the request
to all other replica. All replicas then acknowledge the reception of the request
in an all-to-all communication phase and add the request to their local request vector.
Replicas periodically share their vectors with each other.
The robust function provides (partial) fairness as well.
Note that robustness has also been addressed in other ways, e.g.,
using the leader rotation and a blacklisting mechanism in Spinning \cite{veronese2009spin} or
isolating the incoming traffic of different replicas, and check the performance of the leader in Aardvark \cite{clement2009making}.
\end{tr}

\begin{tr}
\label{tr:fair}
\textbf{Fair.}
This function transforms an unfair protocol, e.g., PBFT, to a fair protocol by adding a preordering phase to the protocol.
In the preordering phase, clients send transactions to all replicas and once a round ends (timer $\tau_5$),
all replicas send a batch of requests in the order received\footnote{The request propagation time can be estimated by measuring network latency or relying on client timestamps for non-Byzantine clients} to the leader.
The leader then initiates consensus on the requests following the order of transactions in received blocks.
Depending on the order-fairness parameter $\gamma$ ($0.5 < \gamma \leq 1$) that defines the fraction of replicas receiving the
transactions in that specific order, at least $4f+1$ replicas ($n > \frac{4f}{2\gamma - 1}$) replicas are needed to provide order fairness \cite{kelkar2020order,kelkar2021themis}
\footnote{Order fairness can be provided using $3f+1$ replicas, however, as shown in \cite{kelkar2021themis},
it either requires a synchronized clock \cite{zhang2020byzantine} or does not provide censorship resistance \cite{kursawe2020wendy}.}.
\end{tr}

\begin{tr}
\label{tr:load}
\textbf{LoadBalancer.}
This function explores a trade-off between communication topology and load balancing where 
load balancing is supported by organizing
replicas in a tree topology, with the leader placed at the root, e.g., Kauri \cite{neiheiser2021kauri}.
The function takes a linear communication phase,
and splits it into $h$ communication phases where $h$ is the height of the tree. 
Each replica then uniformly communicates with its child/parent replicas in the tree.
Using the tree topology, the protocol optimistically assumes all non-leaf replicas are non faulty (assumption $a_3$).
Otherwise the tree needs to be reconfigured (i.e., view change).
\end{tr}
\section{Deriving, Evolving and Inventing Protocols}\label{sec:mapping}

\tikzstyle{protocol} = [rectangle, rounded corners, minimum width=0.6cm, minimum height=0.45cm,text centered, draw=black, fill=blue!10!white]
\tikzstyle{nprotocol} = [rectangle, rounded corners, minimum width=0.6cm, minimum height=0.45cm,text centered, draw=black, fill=lime!30!white]
\tikzstyle{rprotocol} = [rectangle, rounded corners, minimum width=0.6cm, minimum height=0.45cm,text centered, draw=black, fill=red!30!white]

\tikzstyle{arrow} = [->,>=stealth]

\begin{figure}
\tiny
\centering
\begin{tikzpicture}[node distance=2cm]
\node (pbft) [rprotocol] {PBFT \cite{castro2002practical}};
\node (fab) [protocol, below of=pbft] {FaB \cite{martin2006fast}};
\node (bosco) [protocol, left of=fab,xshift=-0.4cm] {Bosco \cite{song2008bosco}};
\node (zyzzyva) [protocol, above of=pbft,xshift=-1.8cm] {Zyzzyva \cite{kotla2007zyzzyva}};
\node (quorum) [protocol, left of=pbft,xshift=-0.5cm,yshift=0.6cm] {Quorum \cite{aublin2015next}};
\node (tendermint) [protocol, above of=pbft] {Tendermint \cite{buchman2018latest}};
\node (lpbft) [protocol, right of=pbft,xshift=0.5cm] {Linear PBFT};
\node (zyzzyva5) [protocol, left of=zyzzyva,xshift=-0.95cm] {Zyzzyva5 \cite{kotla2007zyzzyva}};
\node (q/u) [protocol, left of=quorum,xshift=-0.6cm] {Q/U \cite{abd2005fault}};
\node (hot) [protocol, right of=lpbft,xshift=1.1cm] {HotStuff \cite{yin2019hotstuff}};
\node (sbft) [protocol, above of=lpbft,xshift=1.5cm] {SBFT \cite{gueta2019sbft}};
\node (poe) [protocol, above of=lpbft] {PoE \cite{gupta2021proof}};
\node (kauri) [protocol, above of=hot] {Kauri \cite{neiheiser2021kauri}};
\node (themis) [protocol, below of=hot] {Themis \cite{kelkar2021themis}};
\node (flb) [nprotocol, below of=lpbft] {\flb};
\node (ftb) [nprotocol, right of=flb,xshift=-0.1cm] {\ftb};
\node (prime) [protocol, left of=bosco,xshift=-0.6cm,yshift=1.3cm] {Prime \cite{amir2011prime}};
\node (cheapbft) [protocol, below of=prime,yshift=0.7cm,xshift=0.3cm] {CheapBFT \cite{kapitza2012cheapbft}};

\draw [arrow] (fab) -- node[align=center,yshift=0.15cm] {\ref{tr:resilience}. Resilience} (bosco);
\draw [arrow] (pbft) -- node[align=center,yshift=0.2cm] {\ref{tr:phase}. Linearization} (lpbft);
\draw [arrow] (pbft) -- node[align=center,xshift=0.5cm] {\ref{tr:nodes}. Phase \\reduction} (fab);
\draw [arrow] (pbft) -- node[align=center] {\ref{tr:conflict}. Optimistic \\conflict-free} (quorum);
\draw [arrow] (pbft) -- node[align=center,xshift=-1.1cm] {\ref{tr:active}. Optimistic \\ replica reduction} (cheapbft);
\draw [arrow] (pbft) -- node[align=center] {\ref{tr:view2}. Non-responsive \\ leader rotation} (tendermint);
\draw [arrow] (pbft) -- node[align=right,xshift=-0.7cm] {\ref{tr:2phase}. Speculative \\ execution} (zyzzyva);
\draw [arrow] (quorum) -- node[align=center,yshift=0.15cm] {\ref{tr:resilience}. Resilience} (q/u);
\draw [arrow] (lpbft) -- node[align=center,xshift=0.75cm] {\ref{tr:1phase}. Optimistic  \\phase reduction} (sbft);
\draw [arrow] (lpbft) -- node[align=center,xshift=-0.1cm] {\ref{tr:s1phase}. Speculative  phase \\reduction} (poe);
\draw [arrow] (lpbft) -- node[align=center] {\ref{tr:view1}. Leader \\ Rotation} (hot);
\draw [arrow] (hot) -- node[align=center,xshift=0.4cm] {\ref{tr:load}. Load\\Balancer} (kauri);
\draw [arrow] (flb) -- node[align=center,xshift=0cm] {\ref{tr:load}. Load\\Balancer} (ftb);
\draw [arrow] (hot) -- node[align=center,xshift=0.4cm] {\ref{tr:fair}. Fair} (themis);
\draw [arrow] (zyzzyva) -- node[align=center,yshift=0.15cm] {\ref{tr:resilience}. Resilience} (zyzzyva5);
\draw [arrow] (fab) -- node[align=center,yshift=0.2cm] {\ref{tr:phase}. Linearization} (flb);
\draw [arrow] (lpbft) -- node[align=center,xshift=0.5cm] {\ref{tr:nodes}. Phase \\ reduction} (flb);
\draw [arrow] (pbft) -- node[align=center,xshift=0.1cm,yshift=0.2cm] {\ref{tr:active}. Robust} (prime);

\end{tikzpicture}
\caption{Derivation of protocols from PBFT using design choices}
\label{fig:derivation}
\end{figure}
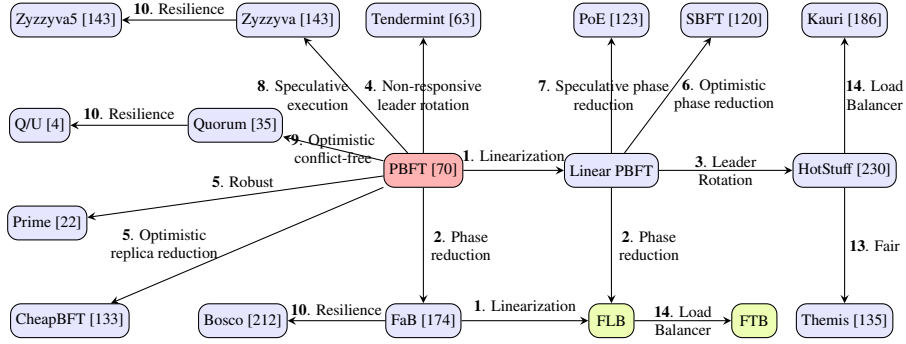

This section demonstrate how \sys is used to derive a wide range of BFT protocols using design choices.
Figure~\ref{fig:derivation} demonstrates the derivation of
a wide spectrum of classical and recent BFT protocols from PBFT using
design choices.
Table~\ref{tbl:protocol} provides insights into how each BFT protocol maps into the \sys design space.
The table also presents the design choices used by each BFT protocol.

\subsection{Case Studies on Protocol Evolution}\label{append:protocols}

In the following case studies, we provide insights into how each BFT protocol maps into the \sys design space, and relate to one another through using design choices. For illustrative purposes, we describe each protocol relative to PBFT along one or more design choices.
Figure~\ref{fig:protocols} focuses on different stages of replicas and demonstrates the communication complexity of
each stage. The figure presents:
(1) the preordering phases used in Themis and Prime,
(2) the three ordering phases,
e.g., {\sf \small pre-prepare}, {\sf \small prepare} or {\sf \small commit} in PBFT (labeled by $o_1$, $o_2$, and $o_3$),
(3) the execution stage,
(4) the view-change stages consisting of {\sf \small view-change} and {\sf \small new-view} phases (labeled by $v_1$ and $v_2$), and
(5) the checkpointing stage.
As can be seen, some protocols do not have all three ordering phases, i.e., using different design choices,
the number of ordering phases is reduced. 
The dashed boxes present the slow-path of protocols, e.g., the third ordering phase of SBFT is used
only in its slow-path. Finally, the order of stages might be changed.
For example, HotStuff runs view-change (leader rotation) for every single message and this leader rotation phase
takes place at the beginning of a consensus instance to synchronize nodes within a view.

Next, we describe each BFT protocol.

\begin{figure}
\begin{minipage}{.195\textwidth}
\vspace{2.5em}
\includegraphics[width= \linewidth]{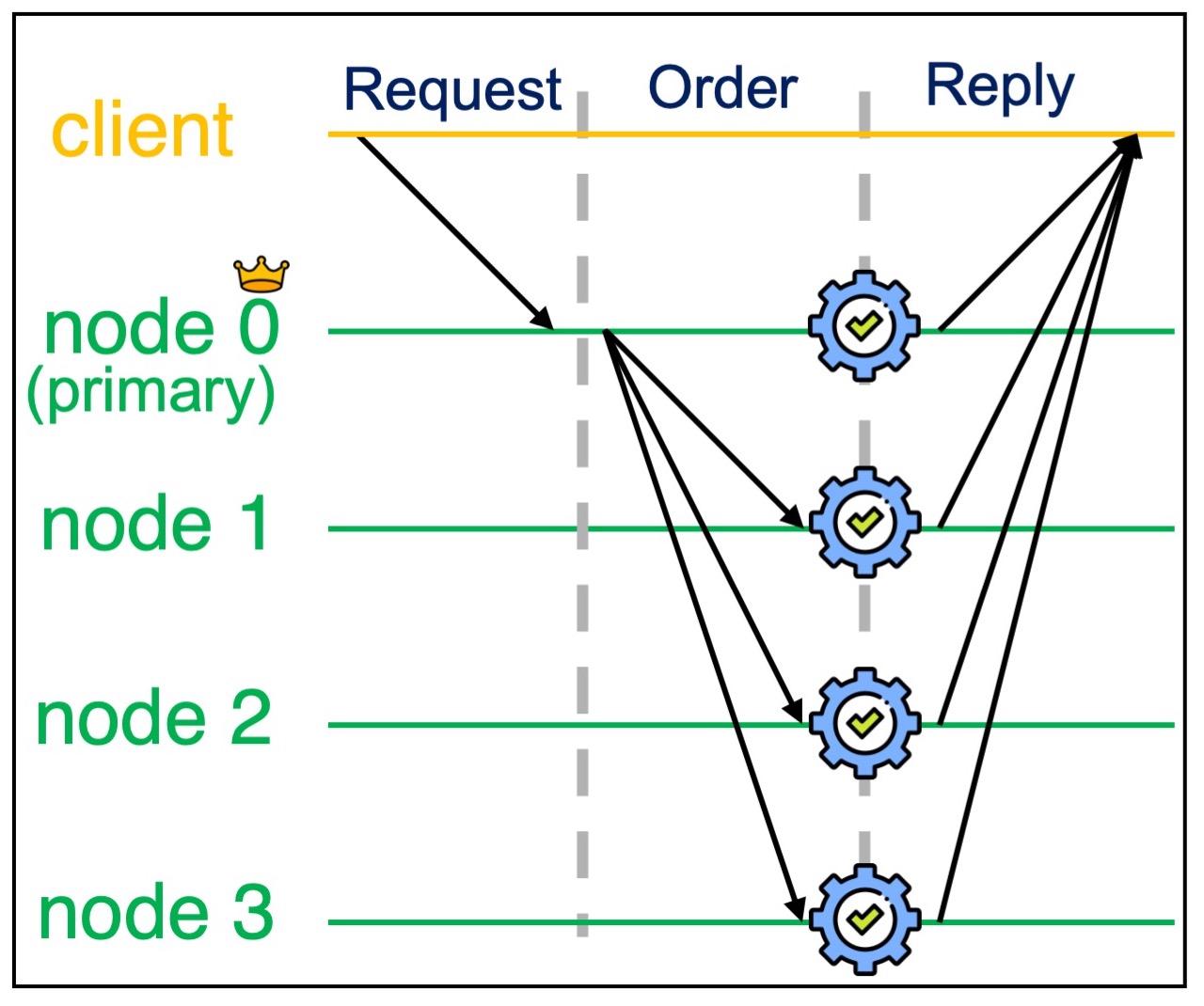}
\caption{Zyzzyva}
\label{fig:zyzzyva1}
\end{minipage}
\begin{minipage}{.285\textwidth}
\vspace{2.5em}
\includegraphics[width= \linewidth]{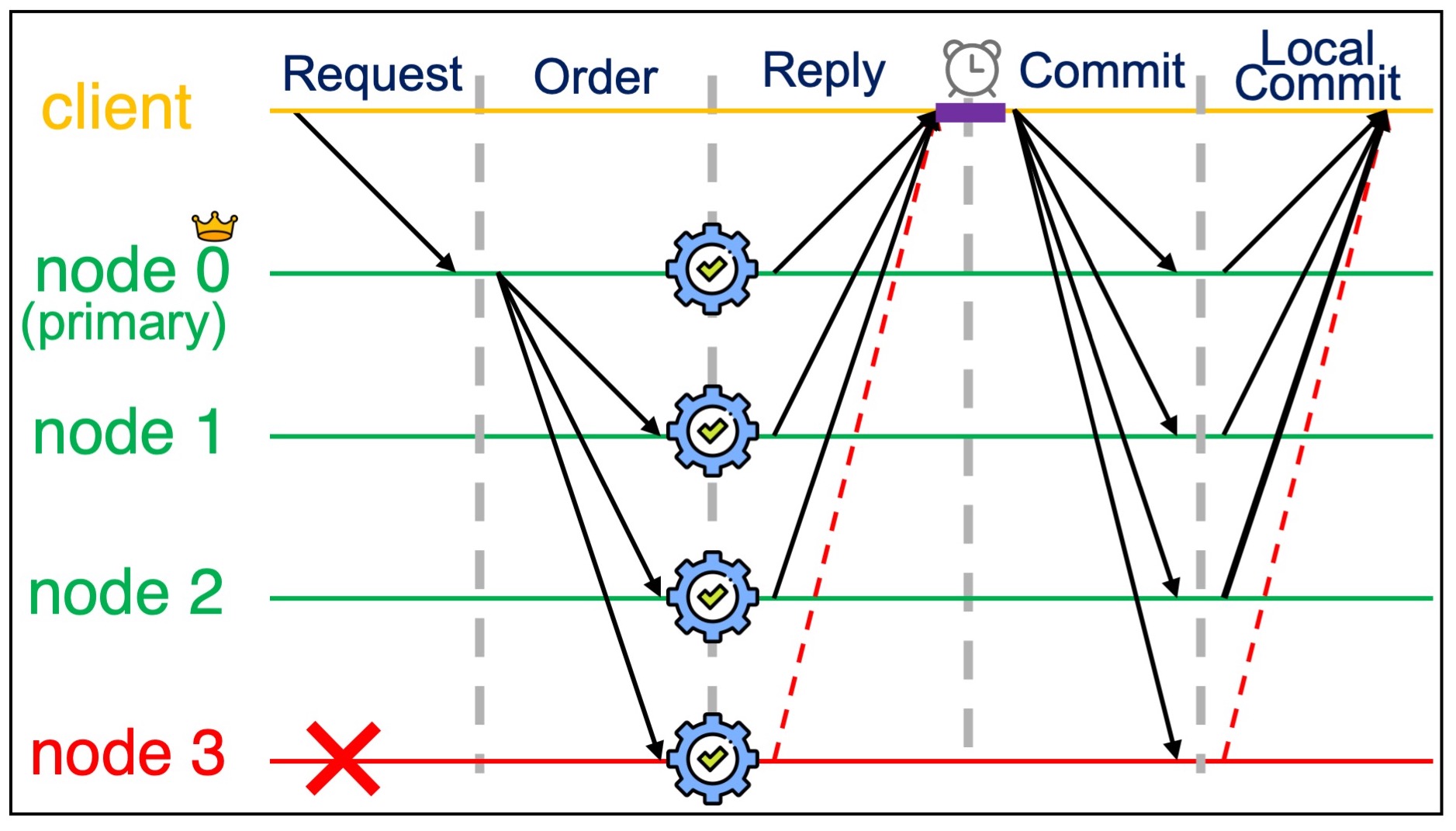}
\caption{Zyzzyva (slow)}
\label{fig:zyzzyva2}
\end{minipage}
\begin{minipage}{.199\textwidth}
\includegraphics[width= \linewidth]{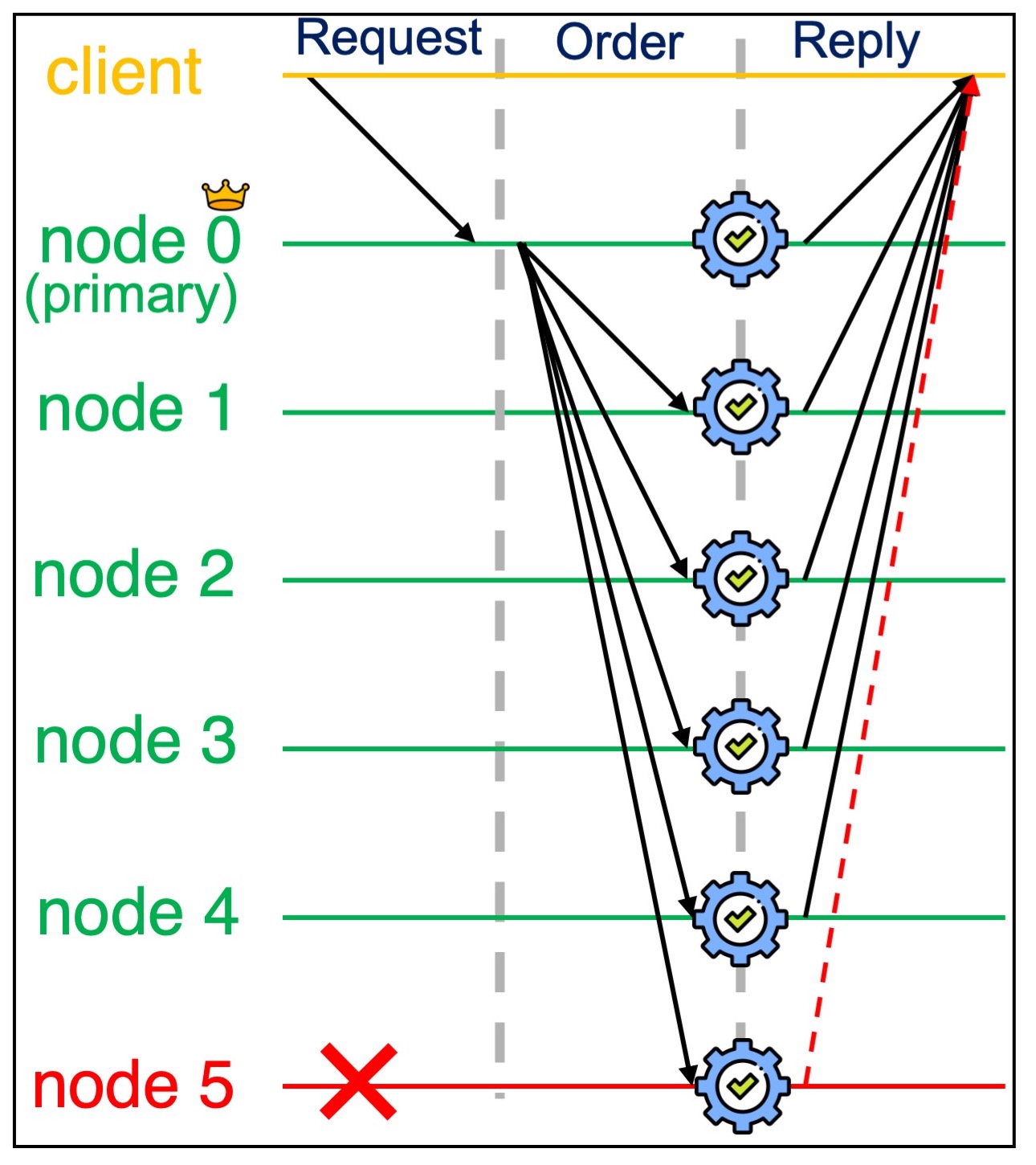}
\caption{Zyzzyva5}
\label{fig:zyzzyva5-1}
\end{minipage}
\begin{minipage}{.286\textwidth}
\includegraphics[width= \linewidth]{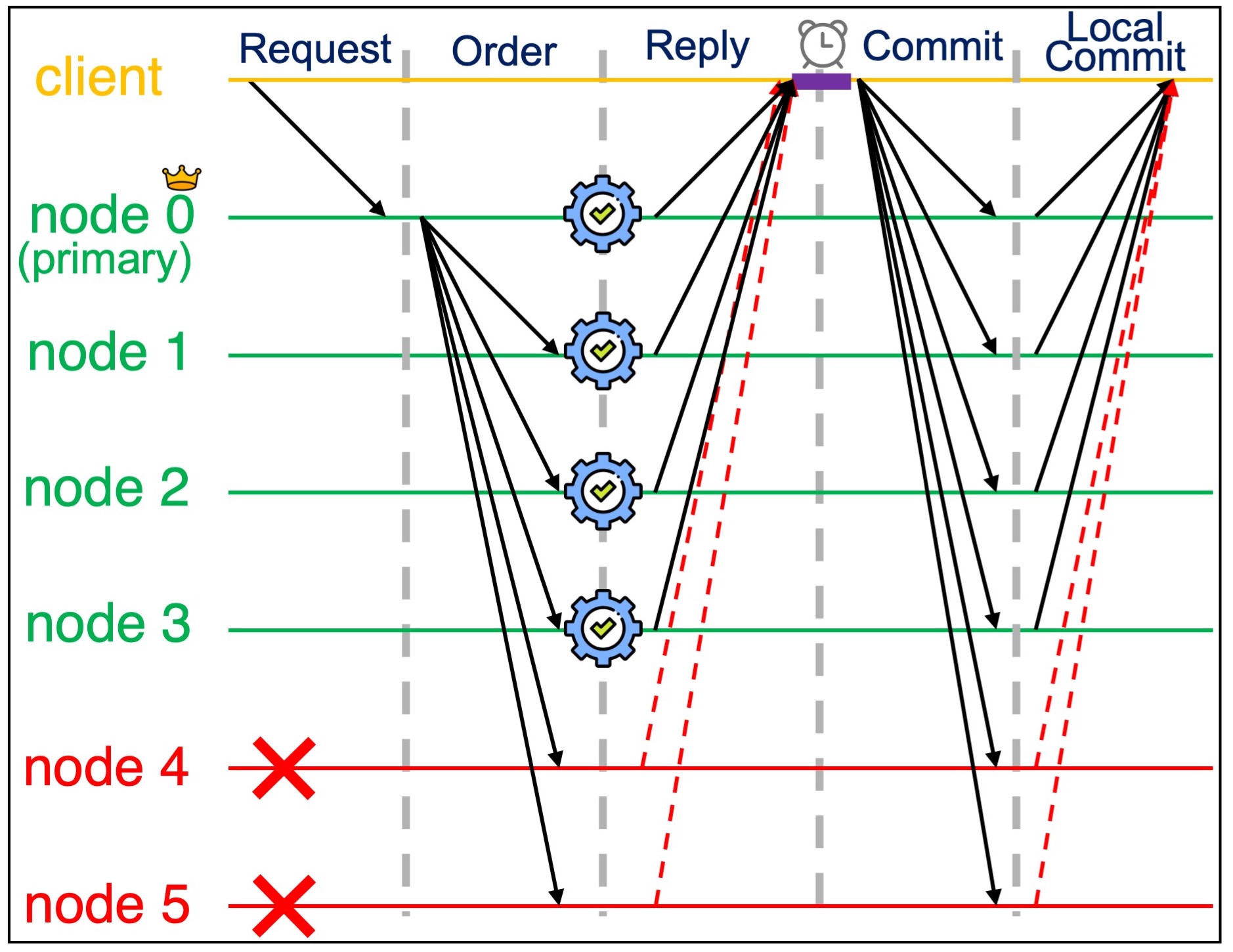}
\caption{Zyzzyva5 (slow)}
\label{fig:zyzzyva5-2}
\end{minipage}
\end{figure}

\begin{figure*}[t]
\begin{minipage}{.437\textwidth}
\includegraphics[width= \linewidth]{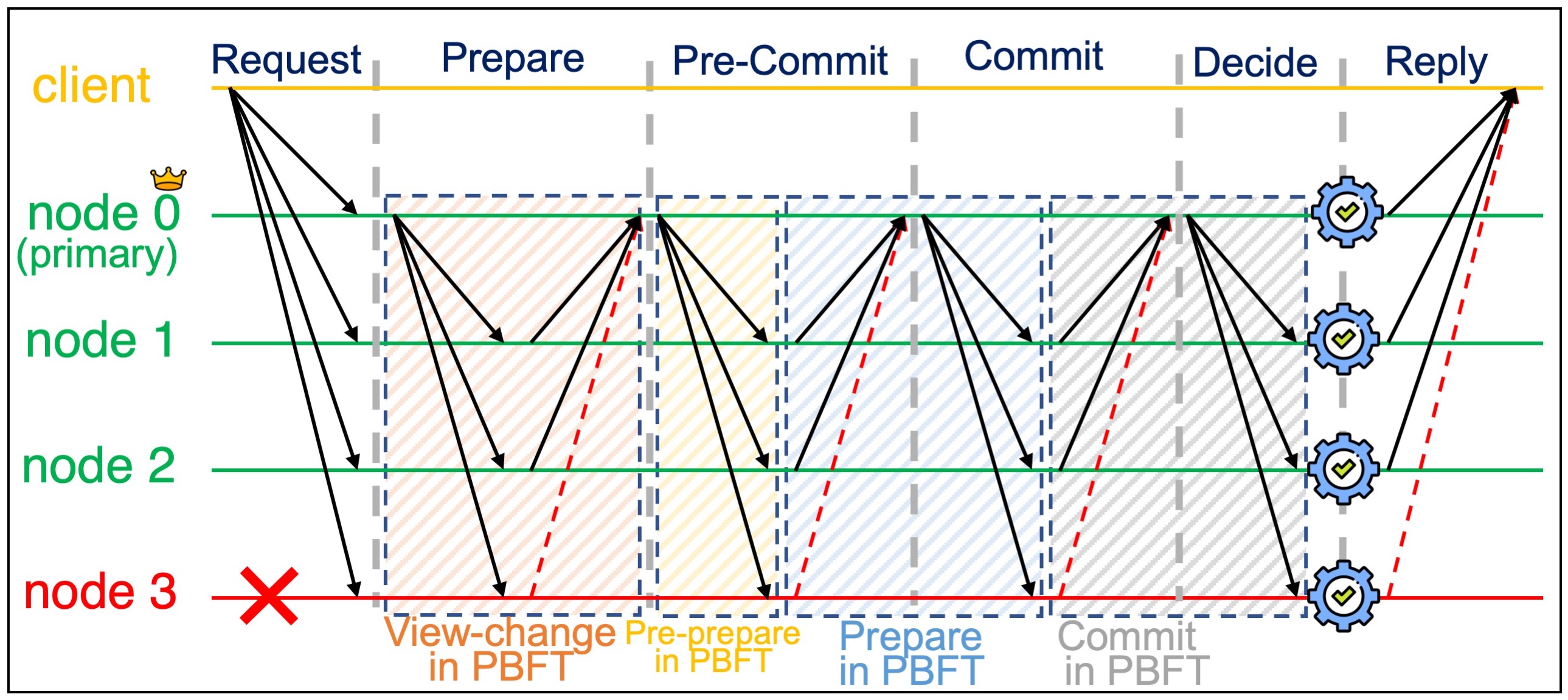}
\caption{HotStuff}
\label{fig:hotstuff}
\end{minipage}
\begin{minipage}{.543\textwidth}
\includegraphics[width= \linewidth]{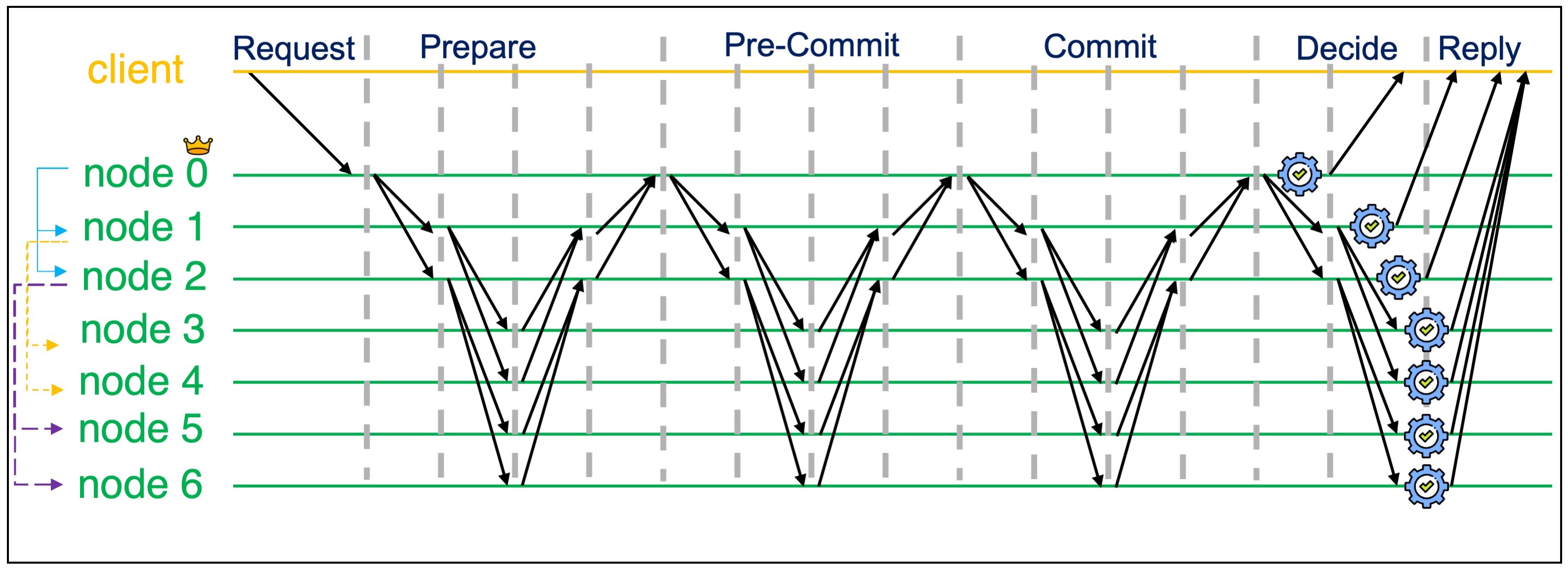}
\caption{Kauri}
\label{fig:kauri}
\end{minipage}
\end{figure*}

\begin{figure*}[t]
\begin{minipage}{.305\textwidth}
\includegraphics[width= \linewidth]{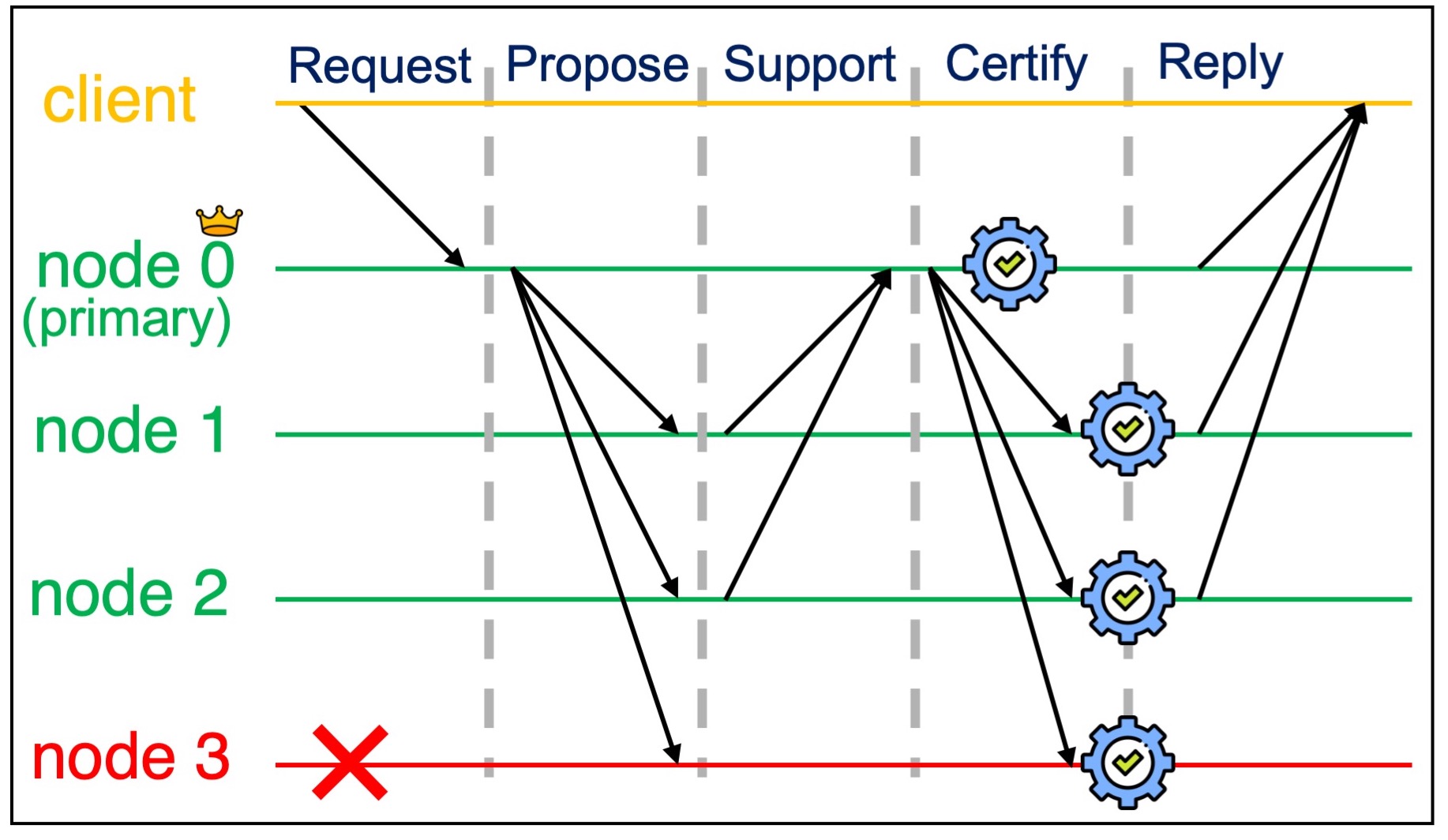}
\caption{PoE}
\label{fig:poe}
\end{minipage}
\begin{minipage}{.315\textwidth}
\includegraphics[width= \linewidth]{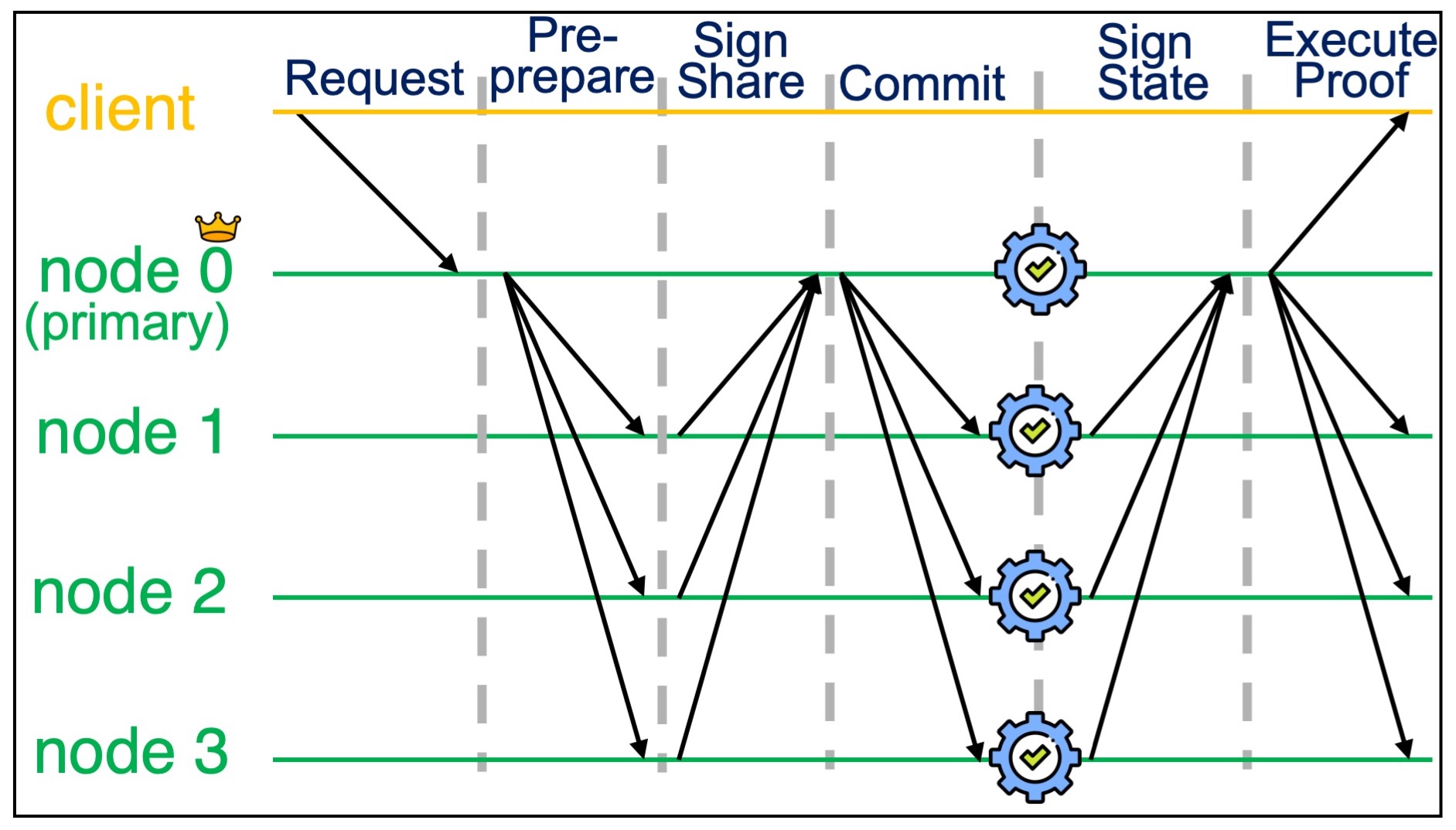}
\caption{SBFT}
\label{fig:sbft}
\end{minipage}
\begin{minipage}{.3668\textwidth}
\includegraphics[width= \linewidth]{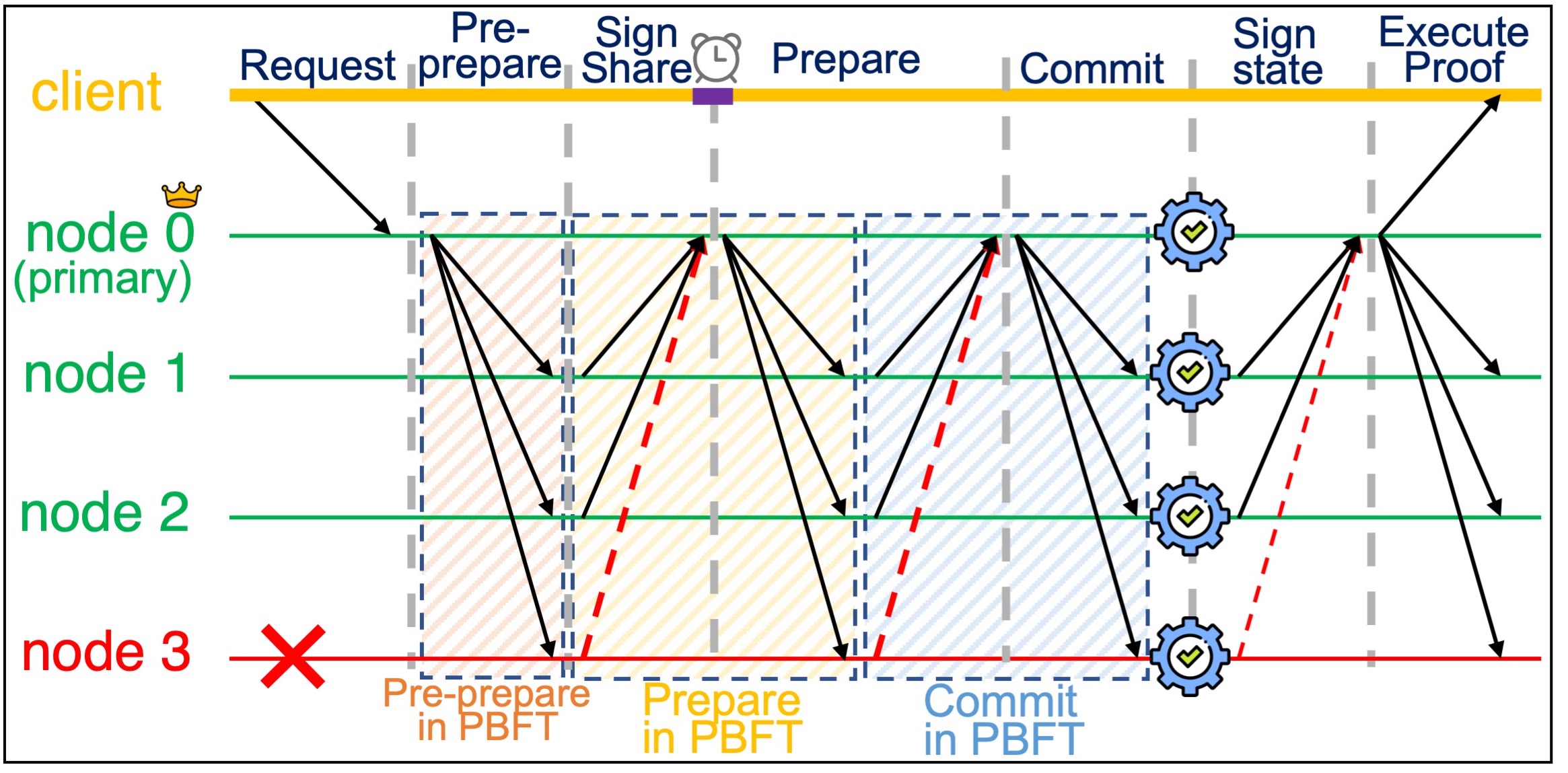}
\caption{SBFT (Linear PBFT)}
\label{fig:sbftlinear}
\end{minipage}
\end{figure*}

\noindent
{\bf Zyzzyva \cite{kotla2007zyzzyva}.}
Zyzzyva\footnote{the view-change stage of the Zyzzyva protocol has a safety violation as described in \cite{abraham2017revisiting}}
(Figure~\ref{fig:zyzzyva1}) can be derived from PBFT using the speculative execution function (design choice \ref{tr:2phase}) of \sys where
assuming the primary and all backups are non-faulty,
replicas speculatively execute requests without running any agreement and send {\sf \small reply} messages to the client.
The client waits for $3f+1$ matching replies to accept the results.
If the timer $\tau_1$ is expired and the client received matching replies from between $2f+1$ and $3f$ replicas,
as presented in Figure~\ref{fig:zyzzyva2},
two more linear rounds of communication is needed to ensure that at least $2f+1$ replicas
have committed the request.
Finally,  Zyzzyva5  is derived from Zyzzyva by using the resilience function (design choice \ref{tr:resilience})
where the number of replicas is increased to $5f+1$ and the protocol is able to tolerate $f$ and $2f$ failures
during its fast and slow path respectively (presented in (Figures~\ref{fig:zyzzyva5-1} and \ref{fig:zyzzyva5-2})
AZyzzyva \cite{aublin2015next,guerraoui2010next} also uses the fast path of Zyzzyva (called ZLight) in its fault-free situations.

\noindent
{\bf PoE \cite{gupta2021proof}.}
PoE Figure~\ref{fig:poe} uses the linearization and speculative phase reduction functions (design choices \ref{tr:phase} and \ref{tr:s1phase}).
PoE does not assumes that all replicas are non-faulty and constructs quorum of $2f+1$ replicas
possibly including Byzantine replicas.
However, since a client waits for $2f+1$ matching reply messages,
all $2f+1$ replicas constructing the quorum need to be well-behaving to guarantee client liveness in the fast path.

\begin{figure*}[t]
\centering
\begin{minipage}{.32\textwidth}
\vspace{2.7em}
\includegraphics[width= \linewidth]{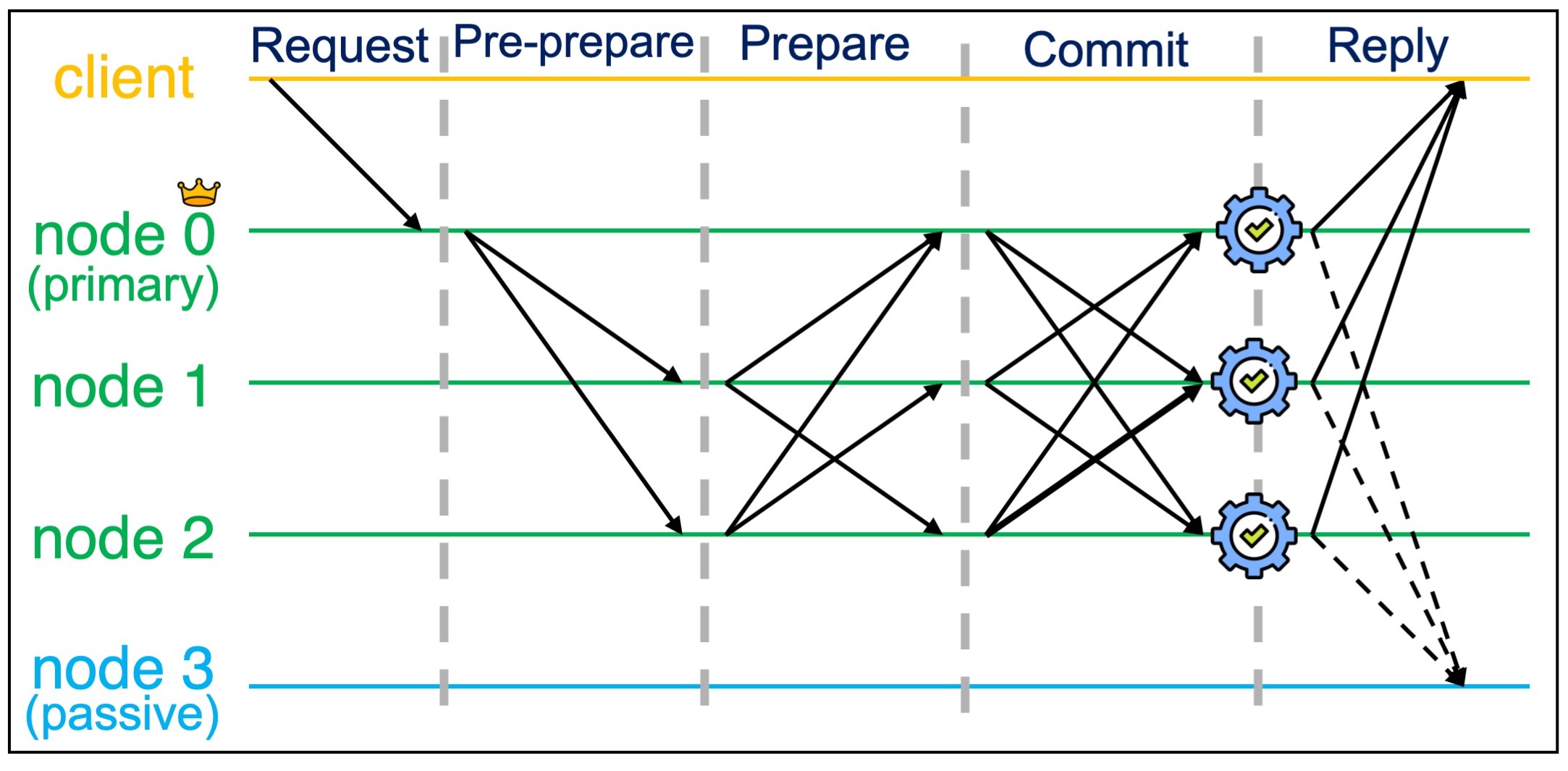}
\caption{CheapBFT}
\label{fig:cheapbft}
\end{minipage}
\begin{minipage}{.26\textwidth}
\includegraphics[width= \linewidth]{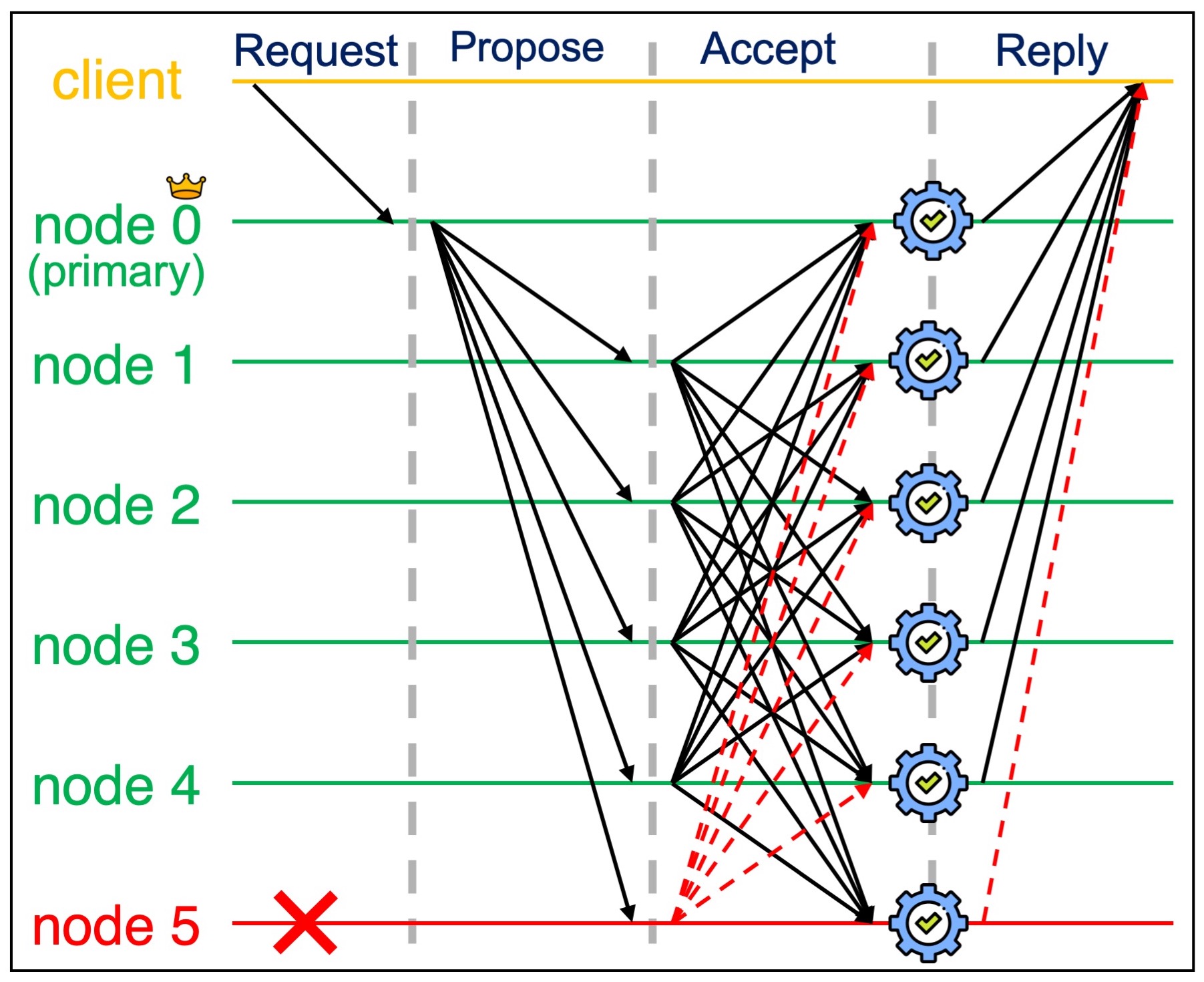}
\caption{FaB}
\label{fig:fab}
\end{minipage}
\begin{minipage}{.149\textwidth}
\includegraphics[width= \linewidth]{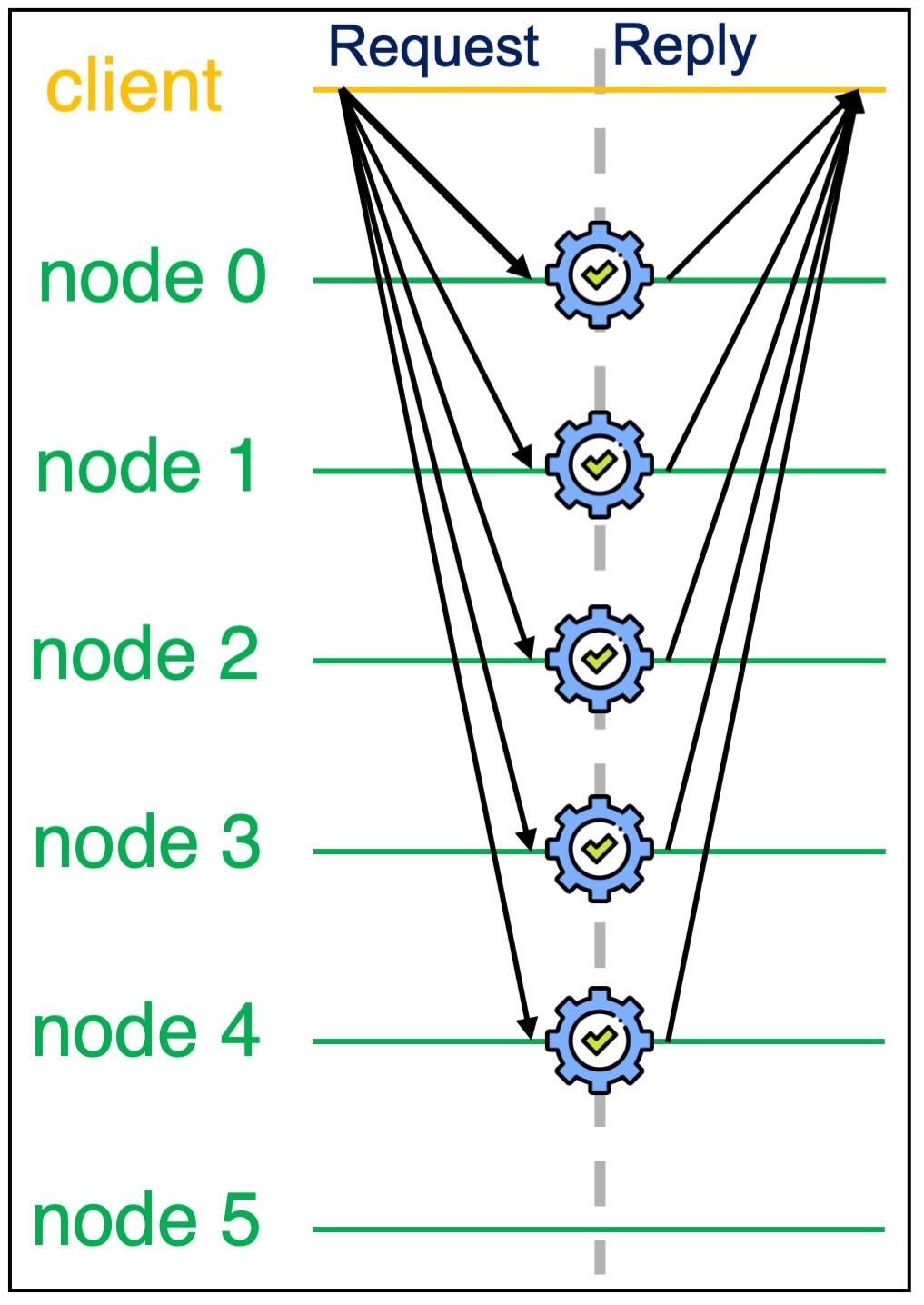}
\caption{Q/U}
\label{fig:qu1}
\end{minipage}
\begin{minipage}{.242\textwidth}
\includegraphics[width= \linewidth]{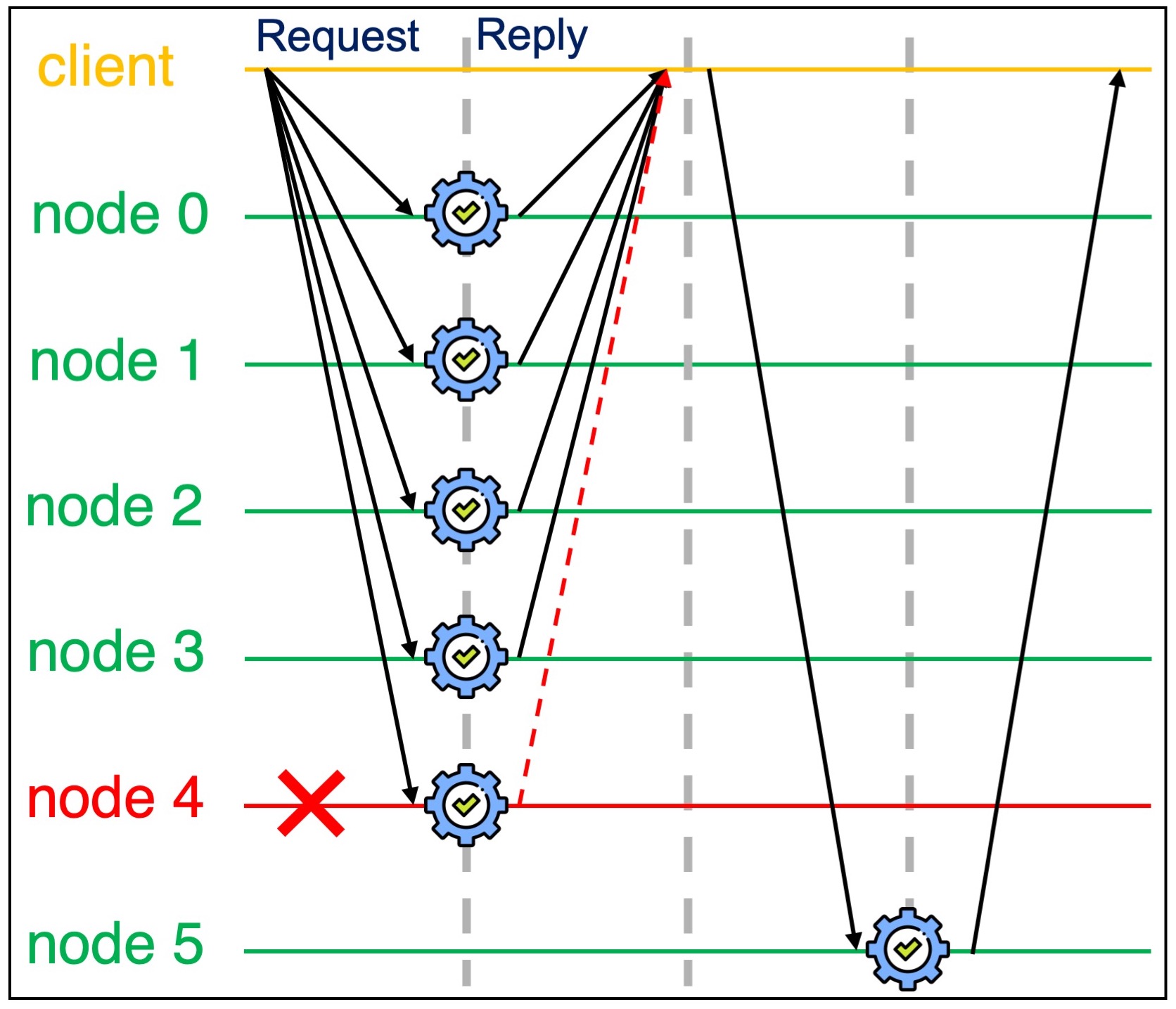}
\caption{Q/U (Normal)}
\label{fig:qu2}
\end{minipage}
\vspace{-1em}
\end{figure*}

\noindent
{\bf SBFT \cite{gueta2019sbft}.}
\sys derives SBFT\footnote{SBFT tolerates both crash and Byzantine failure
($n=3f+2c+1$ where $c$ is the number of crashed replicas).Since the focus of this paper is on Byzantine failures, we consider a variation of SBFT where $c=0$.} from PBFT using the linearization and optimistic phase reduction functions
(design choices \ref{tr:phase} and \ref{tr:1phase}).
SBFT presents an optimistic fast path (Figure~\ref{fig:sbft}) assuming all replicas are non-faulty.
If the primary does not receive messages from {\em all} backups (in the \two phase) and its timer is expired
(i.e., non-responsiveness timer $\tau_3$),
SBFT switches to its slow path (Figure~\ref{fig:sbftlinear}) and requires two more linear rounds of communication (\three phase).
The Twin-path nature of SBFT requires replicas to sign each message with two schemes (i.e., $2f+1$ and $3f+1$).
To send replies to the client,
a single (collector) replica receives replies from all replicas and sends a single (threshold) signed reply message.

\noindent
{\bf HotStuff \cite{yin2019hotstuff}.}
HotStuff (Figure~\ref{fig:hotstuff}) can be derived from PBFT using the linearization and leader rotation functions
(design choices \ref{tr:phase} and \ref{tr:view1}) of \sys.
The Chained HotStuff (performance optimization \ref{optim:pipe}) benefits from pipelining to reduce the latency of request processing.

\noindent
{\bf Tendermint \cite{kwon2014tendermint,buchman2016tendermint,buchman2018latest}.}
Tendermint\footnote{Tendermint uses a Proof-of-Stake variation of PBFT
where each replica has a voting power equal to its stake (i.e., locked coins).} leverages the non-responsive leader rotation function (design choice \ref{tr:view2})
to rotate leader without adding any new phase. The new leader, however, needs to wait
for a predefined time (timer $\tau_4$), i.e.,
the worst-case time it takes to propagate messages over a wide-area peer-to-peer gossip network, before proposing a new block.
Tendermint also uses timers in all phases where a replica discard the request if it does not receive $2f+1$ messages
before the timeout (timer $\tau_6$).
Note that the original Tendermint uses a gossip all-to-all mechanism and has $\mathcal{O}(n\log n)$ message complexity.

\noindent
{\bf Themis \cite{kelkar2021themis}.}
Themis is derived from HotStuff using the fair function (design choice \ref{tr:fair}).
Themis add a new all-to-all preordering phase where
replicas send a batch of requests in the order they received to the leader replica and
the leader  propose requests in the order received (depending on the order-fairness parameter $\gamma$) \cite{kelkar2020order}.
Themis requires at least $4f+1$ replicas (if $\gamma =1$) to provide order fairness.

\noindent
{\bf Kauri \cite{neiheiser2021kauri}.}
Kauri (Figure~\ref{fig:kauri}) can be derived from HotStuff using  the loadbalancer function (design choice \ref{tr:load}) that
maps the star topology to the tree topology.
The height of the tree is $h= \log_d n$ where $d$ is the fanout of each replica.

\noindent
{\bf CheapBFT \cite{kapitza2012cheapbft}.}
CheapBFT  (Figure~\ref{fig:cheapbft}) and its revised version, \textsc{ReBFT}\cite{distler2016resource} is derived from PBFT using
the optimistic replica reduction function (design choice \ref{tr:active}).
Using trusted hardware (performance optimization O\ref{optim:trusted}), a variation of \textsc{ReBFT}, called \textsc{RwMinBFT}, processes
requests with $f+1$ active and $f$ passive replicas in its normal case (optimistic) execution.

\noindent
{\bf FaB \cite{martin2006fast}.}
FaB\footnote{FaB similar to a family of Paxos-like protocol
separates proposers from acceptors.
In our implementation of FaB, however, replicas act as both proposers and acceptors.} 
(Figure~\ref{fig:fab}) uses the phase reduction function (design choice \ref{tr:nodes})
to reduce one phase of communication while requiring $5f+1$ replicas.
Fab does not use authentication in its ordering stage, however,
requires signatures for the view-change stage (design choice \ref{tr:auth}).
Note that using authentication, $5f-1$ replicas is sufficient to reduce one phase of communication \cite{kuznetsov2021revisiting,abraham2021good}.

\noindent
{\bf Prime \cite{amir2011prime}.}
Prime is derived from PBFT using the robust functions (design choice \ref{tr:robust}).
In prime a preordering stage is added where replicas exchange the requests they receive from clients and
periodically share a vector of all received requests, which they expect the leader to order request following those vectors.
In this way, replicas can also monitor the leader to order requests in a fair manner.

\noindent
{\bf Q/U \cite{abd2005fault}.}
Q/U (Figure~\ref{fig:qu1}) utilizes optimistic conflict-free and resilience functions (design choices \ref{tr:conflict} and \ref{tr:resilience}).
Clients play the proposer role and replicas immediately execute an update request
if the object has not been modified since the client's last query.
Since Q/U is able to tolerate $f$ faulty replicas,
a client can optionally communicate with a subset ($4f+1$) of replicas (preferred quorum).
The client communicate with additional replicas only if it does not received \reply from all replicas of the preferred quorum (Figure~\ref{fig:qu2}).
Both signatures (for large $n$) and MACs (for small $n$) can be used for authentication in Q/U.
Quorum \cite{aublin2015next} uses a similar technique with $3f+1$ replicas, i.e., 
only the conflict-free function (design choices \ref{tr:conflict}) has been used.

\subsection{Deriving Novel BFT Protocols}

\begin{figure*}[t]
\centering
\begin{minipage}{.317\textwidth}
\includegraphics[width= \linewidth]{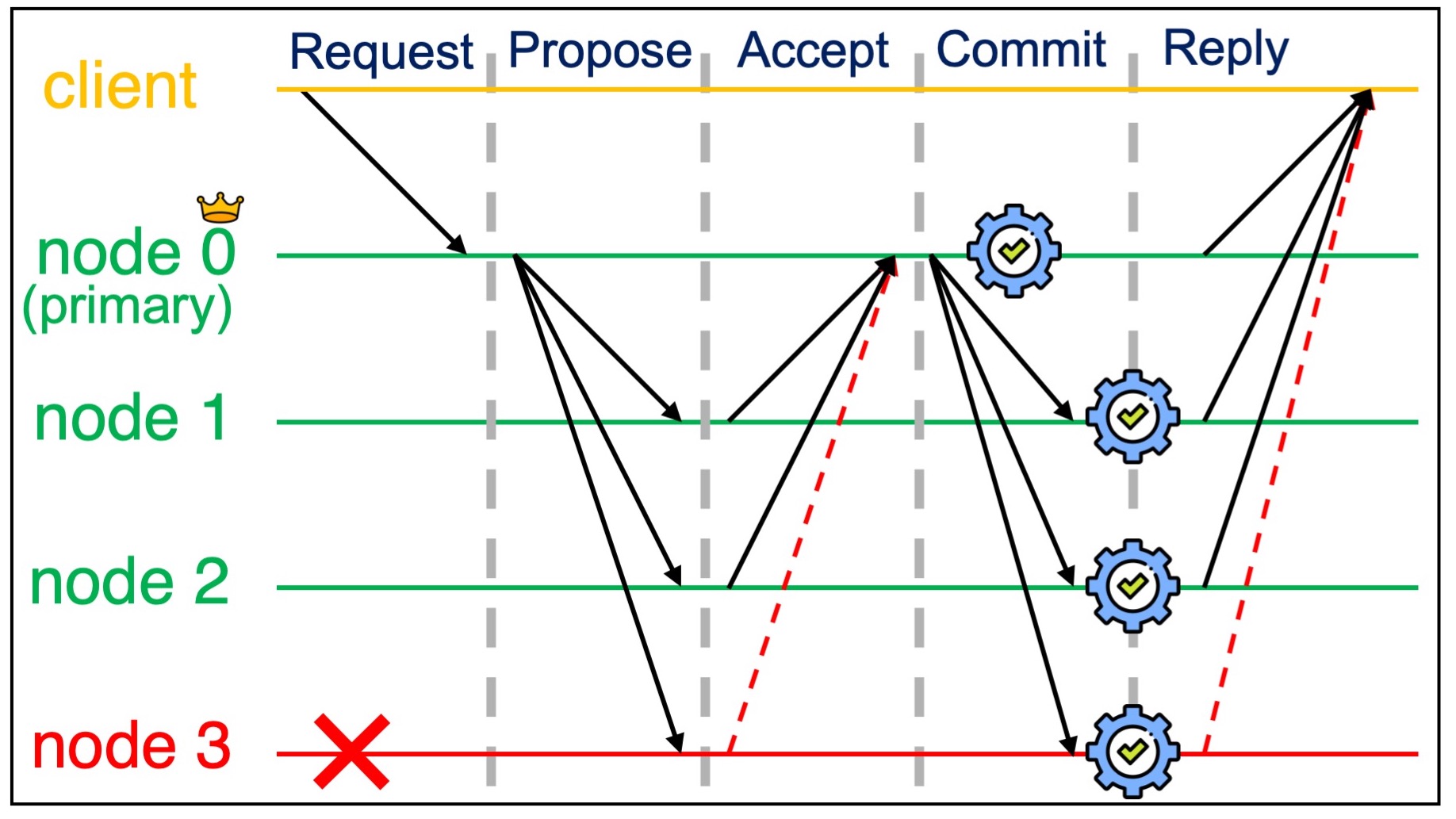}
\caption{\flb}
\label{fig:flb}
\end{minipage}\hspace{2em}
\begin{minipage}{.319\textwidth}
\includegraphics[width= \linewidth]{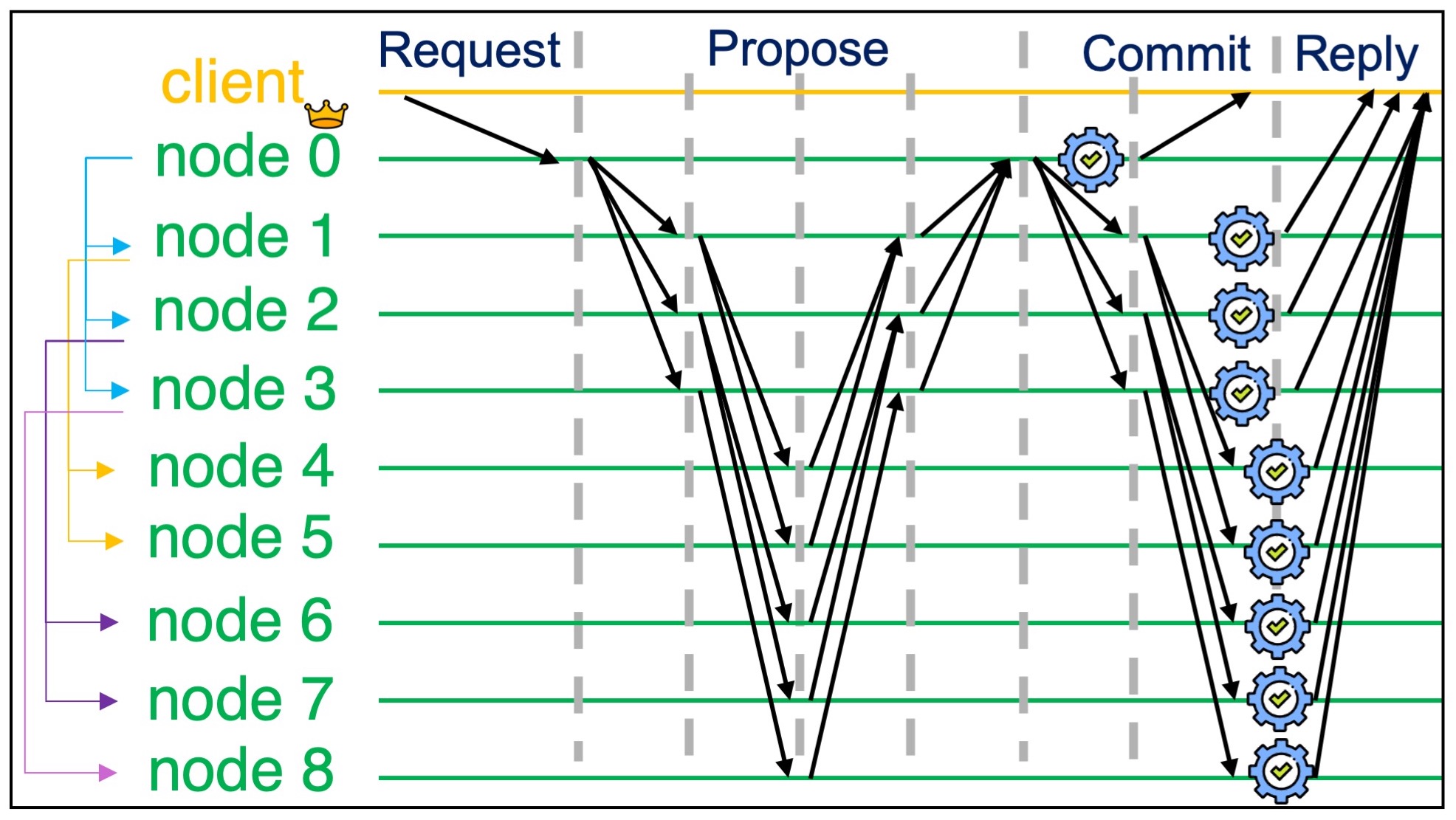}
\caption{\ftb}
\label{fig:ftb}
\end{minipage}
\end{figure*}

The previous case studies demonstrate the value of \sys in providing a unified platform for analyzing
the strengths and weaknesses of a range of existing BFT protocols.
\sys's utility goes beyond an experimental platform towards a {\em discovery} tool as well.
The system provides a systematic way to explore new valid points in the design space
and help BFT researchers uncover {\em novel} BFT protocols.
We uncover several such new protocols, although not all are necessarily practical or interesting.
For example, simply making a protocol fair by adding the preordering phase of fairness
results in a new protocol.
While this is an interesting insight, the resulting protocol may have limited practical impact.
We select as highlights two new BFT protocols (\flb and \ftb)
that are new and have practical value that we have uncovered using \sys.

\noindent
{\bf Fast Linear BFT (FLB).}
{\em \flb} is a fast linear BFT protocol that commits transactions
in two phases of communication with linear message complexity.
To achieve this, \flb uses the linearization and phase reduction through redundancy functions
(design choices \ref{tr:phase} and \ref{tr:nodes}).
\flb requires $5f-1$ replicas (following the lower bound results on fast Byzantine agreement \cite{kuznetsov2021revisiting,abraham2021good}).
The ordering stage of \flb is similar to the fast path of SBFT
in terms of the linearity of communication and the number of phases.
However, \flb expands the network size to tolerate $f$ failures
(in contrast to SBFT, which optimistically assumes all replicas are non-faulty).

\noindent
{\bf Fast Tree-based balanced BFT (FTB).}
A performance bottleneck of consensus protocols is the computing and bandwidth capacity of the leader.
While Kauri \cite{neiheiser2021kauri} leverages a tree communication topology (design choice \ref{tr:load})
to distribute the load among all replicas,
Kauri requires $7h$ phases of communication to commit each request, where $h$ is the height of the communication tree.
\ftb reduces the latency of Kauri based on two observations.
First, we noticed that while Kauri is implemented on top of HotStuff, it does not use the leader rotation mechanism.
As a result, it does not need the two linear phases of HotStuff ($2h$ phases in Kauri)
that are added for the purpose of leader rotation (design choice \ref{tr:view1}).
Second, similar to \flb, we can use the phase reduction through redundancy function (design choice \ref{tr:nodes})
to further reduce $2h$ more phases of communication.
\ftb establishes agreement with $5f-1$ replicas in $3h$ phases.
\ftb also uses the pipelining stretch mechanism of Kauri, where
the leader continuously initiates consensus instances before receiving a response from its child nodes for the first instance.
This is similar to the out-of-order processing mechanism (performance optimization O \ref{optim:order})
used by many BFT protocols.

\section{Implementation}\label{sec:implementation}

\sys enables users, e.g., application developers, to analyze and navigate the BFT landscape by 
querying their required BFT protocol characteristics.
Using a {\em constraint checker},
\sys finds all plausible points (i.e., BFT protocols) in the design space that satisfy the input query.
Moreover, \sys provides an {\em execution engine} that enables users to implement different BFT protocols.

\begin{figure}[t]
\centering
\includegraphics[width= 0.6 \linewidth]{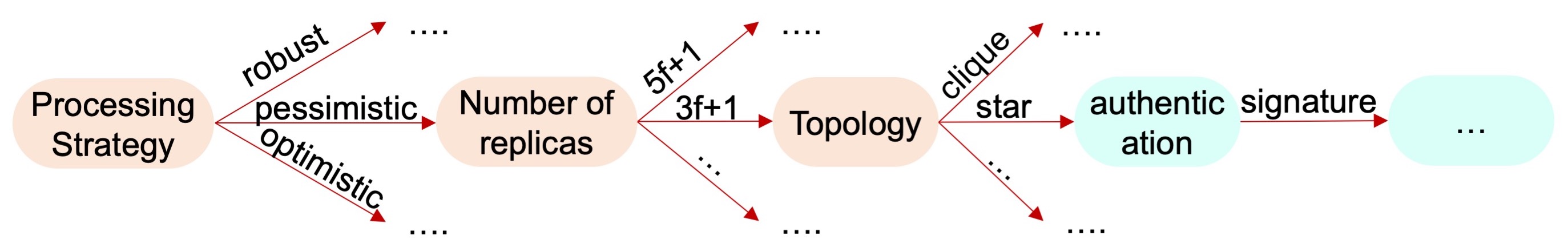}
\caption{Part of a decision tree constructed by the constraint checker}
\label{fig:tree}
\end{figure}

\noindent
{\bf Constraint checker}. The constraint checker uses a decision tree-like algorithm to
find all plausible points in the design space where each node of the tree is labeled with a dimension, and
outgoing edges represent possible values for that particular dimension.
A user issues a query that indicates, for each dimension,
an assigned value (chosen from a preset of values) that characterizes the requirements of their application.
The user can customize some dimensions while leaving the rest unspecified.
For example the user might search for a {\sf pessimistic} BFT protocol
with {\sf 3f+1} nodes and {\sf linear} message complexity.
Based on the initial query, the constraint checker begins with the dimensions that the end-user has specified, e.g.,
processing strategy, the number of replicas, and topology
in Figure~\ref{fig:tree},
and then checks plausible values for unspecified dimensions, e.g., since the end-user has chosen the star topology,
the constraint checker chooses signatures for authentication to be able to provide non-repudiation.
The leaf nodes of the decision tree specify the candidate BFT protocols.
If some queries do not match any points in the design space (e.g. an impossible protocol such as a {\sf pessimistic} {\sf linear} BFT protocol with {\sf 2} communication phases and {\sf 3f+1} replicas), the empty set is returned to the user.

\noindent
{\bf Execution engine}.
The user receives a list of BFT protocols, e.g.,
HotStuff \cite{yin2019hotstuff} is returned as a {\sf pessimistic} {\sf linear} protocol with {\sf 3f+1} nodes.
\sys maintains a library including the implementation of common BFT protocols.
When a protocol is chosen by the user, \sys simply returns it
if the protocol implementation already exists.
Otherwise, the system generates a protocol {\small\tt config} file for the protocol,
including all dimensions and the chosen value for each dimension.
The {\small\tt config} file is then used by \sys to implement the {\small\tt stateMachine} of the protocol.
The {\small\tt stateMachine} includes different states and their transitions for each role (leader, backups, and clients),
exchange messages, quorum conditions, etc.
The execution engine takes the {\small\tt stateMachine} and
a set of related classes that are defined outside the {\small\tt stateMachine}
and handles environmental conditions such as message exchanges, timers, clocks, message validation, etc.
to implement the protocol.
Once the protocol implementation is completed,
the execution engine initializes replicas and clients (based on the system config), deploys the {\small\tt stateMachine} on
nodes, coordinates key exchange, etc.
At this point, the system is ready to run experiments.

\sys is implemented in Java. The modular design of \sys enables a fair and efficient
evaluation of BFT protocols using identical libraries, cryptographic functions, etc.
In particular, we use {\small \tt java.security} and {\small \tt javax.crypto} libraries for cryptographic operations,
the {\small \tt eo-YAML} library to facilitate the storage and retrieval of  BFT protocol configuration files, protocol buffers ({\small \tt com.google.protobuf}) to serialize data, and {\small \tt Java streams} for parallel processing.
\section{Experimental Evaluation}\label{sec:eval}

Our evaluation studies the practical impact of the design dimensions and the exposed trade-offs presented as design choices
on the performance of BFT protocols to reveal the strengths and weaknesses of existing BFT protocols.
The goal is to test the capability of the \sys design space to analyze the performance of different protocols
that were proposed in diverse settings and different contexts under one unified framework.
We evaluate the performance of BFT protocols in 
typical experimental scenarios used for existing BFT protocols and permissioned blockchains, including
(1) varying the number of replicas,
(2) under a backup failure,
(3) multiple request batch sizes, and
(4) a geo-distributed setup.

All protocols listed in Table~\ref{tbl:protocol} are implemented in \sys.
Using the platform, we also experimented with many new protocols resulting from the combination of design choices.
Due to space limitations, we present the performance evaluation of a subset of protocols.
In particular, we evaluate PBFT, Zyzzyva, SBFT, FaB, PoE, (Chained) HotStuff, Kauri, Themis, and
two of the more interesting new variants (\flb and \ftb).
This set of protocols enables us to see the impact of
design choices~\ref{tr:phase}, \ref{tr:nodes}, \ref{tr:view1}, \ref{tr:1phase}, \ref{tr:s1phase},
\ref{tr:2phase}, \ref{tr:resilience}, \ref{tr:auth}, \ref{tr:fair}, and \ref{tr:load} (discussed in Section~\ref{sec:design}).
We also use the out-of-order processing technique (optimization O\ref{optim:order})
for protocols with a stable leader and
the request pipelining technique (optimization O\ref{optim:pipe}) 
for protocols with a rotating leader.
In our experiments, Kauri and \ftb are deployed on trees of height $2$ and
the order-fairness parameter $\gamma$ of Themis is considered to be $1$ (i.e., $n = 4f+1$).
Kauri does not use the rotating leader mechanism (although it was developed as an extension of HotStuff).
We use $4$ as the base pipelining stretch
for both Kauri and \ftb and
change it depending on the batch size and deployment setting (local vs. geo-distributed).

\begin{figure}[t]
\centering
\begin{minipage}{.33\textwidth}\centering
\vspace{0.9em}
\begin{tikzpicture}[scale=0.7]
\begin{axis}[
    xlabel={Number of replicas},
    ylabel={Throughput [ktrans/sec]},
    xmin=0, xmax=105,
    ymin=0, ymax=260,
    xtick={4,16,32,64,100},
    ytick={50,100,150,200,250},
    legend columns=10,
    legend style={at={(axis cs:4,-45)},anchor=north west}, 
    ymajorgrids=true,
    grid style=dashed,
]

\addplot[
    color=violet,
    mark=*,
    mark size=3pt,
    line width=0.2mm,
    ]
    coordinates {
    (4,148)(16,116)(31,95)(64,72)(91,57)(100,51)};

\addplot[
    color=cyan,
    mark=pentagon*,
    mark size=3pt,
    line width=0.2mm,
    ]
    coordinates {
    (4,250)(16,218)(31,189)(64,163)(91,141)(100,138)};
    
\addplot[
    color=magenta,
    mark=diamond,
    mark size=3pt,
    line width=0.2mm,
    ]
    coordinates {
    (4,190)(16,167)(31,143)(64,106)(91,89)(100,85)};
    
\addplot[
    color=orange,
    mark=diamond*,
    mark size=3pt,
    line width=0.2mm,
    ]
    coordinates {
    (4,194)(16,175)(31,153)(64,136)(91,124)(100,119)};

\addplot[
    color=black,
    mark=pentagon,
    mark size=3pt,
    line width=0.2mm,
    ]
    coordinates {
    (6,175)(16,147)(31,121)(66,91)(91,68)(101,65)};
    
\addplot[
    color=purple,
    mark=o,
    mark size=3pt,
    line width=0.2mm,
    ]
    coordinates {
    (4,170)(16,154)(31,131)(64,107)(91,96)(100,94)};
    
 \addplot[
    color=red,
    mark=square,
    mark size=3pt,
    line width=0.2mm,
    ]
    coordinates {
    (4,142)(16,121)(31,109)(64,98)(91,89)(100,86)};

\addplot[
    color=blue,
    mark=triangle*,
    mark size=3pt,
    line width=0.2mm,
    ]
    coordinates {
    (5,28)(17,21)(33,19)(65,17)(93,14)(101,13)};
    
 \addplot[
    color=teal,
    mark=otimes,
    mark size=3pt,
    line width=0.2mm,
    ]
    coordinates {
    (4,194)(14,174)(29,158)(64,132)(89,119)(99,115)};
    
 \addplot[
    color=gray,
    mark=square*,
    mark size=3pt,
    line width=0.2mm,
    ]
    coordinates {
    (4,177)(14,167)(29,155)(64,133)(89,125)(99,121)};

\addlegendentry{PBFT}
\addlegendentry{Zyzzyva}
\addlegendentry{SBFT}
\addlegendentry{PoE}
\addlegendentry{FaB}
\addlegendentry{HotStuff}
\addlegendentry{Kuari}
\addlegendentry{Themis}
\addlegendentry{\flb}
\addlegendentry{\ftb}
 
\end{axis}
\end{tikzpicture}
\end{minipage}\hspace{4em}
\begin{minipage}{.33\textwidth} \centering
\begin{tikzpicture}[scale=0.7]
\begin{axis}[
    xlabel={Number of replicas},
    ylabel={Latency [ms]},
    xmin=0, xmax=105,
    ymin=0, ymax=150,
    xtick={4,16,32,64,100},
    ytick={0,30,60,90,120}, 
    ymajorgrids=true,
    grid style=dashed,
]

\addplot[
    color=violet,
    mark=*,
    mark size=3pt,
    line width=0.2mm,
    ]
    coordinates {
    (4,3)(16,7)(31,11)(64,28)(91,47)(100,56)};

\addplot[
    color=cyan,
    mark=pentagon*,
    mark size=3pt,
    line width=0.2mm,
    ]
    coordinates {
    (4,3)(16,4)(31,10)(64,23)(91,41)(100,51)};
    
\addplot[
    color=magenta,
    mark=diamond,
    mark size=3pt,
    line width=0.2mm,
    ]
    coordinates {
    (4,3)(16,6)(31,10)(64,24)(91,30)(100,35)};
    
\addplot[
    color=orange,
    mark=diamond*,
    mark size=3pt,
    line width=0.2mm,
    ]
    coordinates {
    (4,3)(16,6)(31,9)(64,16)(91,23)(100,26)};
    
\addplot[
    color=black,
    mark=pentagon,
    mark size=3pt,
    line width=0.2mm,
    ]
    coordinates {
    (6,3)(16,5)(31,8)(66,21)(91,32)(101,38)};
    
\addplot[
    color=purple,
    mark=o,
    mark size=3pt,
    line width=0.2mm,
    ]
    coordinates {
    (4,5)(16,11)(31,17)(64,37)(91,64)(100,76)};
    
 \addplot[
    color=red,
    mark=square,
    mark size=3pt,
    line width=0.2mm,
    ]
    coordinates {
    (4,8)(16,18)(31,31)(64,59)(91,99)(100,117)};

\addplot[
    color=blue,
    mark=triangle*,
    mark size=3pt,
    line width=0.2mm,
    ]
    coordinates {
    (5,13)(17,18)(33,27)(65,62)(93,113)(101,137)};
    
 \addplot[
    color=teal,
    mark=otimes,
    mark size=3pt,
    line width=0.2mm,
    ]
    coordinates {
    (4,3)(14,5)(29,9)(64,19)(89,27)(99,31)};
    
 \addplot[
    color=gray,
    mark=square*,
    mark size=3pt,
    line width=0.2mm,
    ]
    coordinates {
    (4,6)(14,12)(29,20)(64,36)(89,54)(99,63)};

\end{axis}
\end{tikzpicture}
\end{minipage}
\caption{Performance with different number of replicas}
  \label{fig:nodes}
\end{figure}
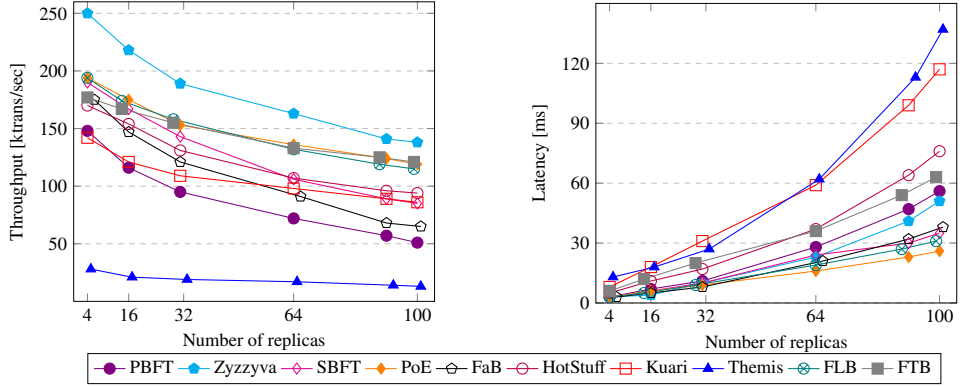

The experiments were conducted on the Amazon EC2 platform.
Each VM is a c4.2xlarge instance with 8 vCPUs and 15GB RAM,
Intel Xeon E5-2666 v3 processor clocked at 3.50 GHz.
When reporting throughput,
we use an increasing number of client requests until the end-to-end throughput is saturated
and state the throughput and latency just below saturation.
The results reflect end-to-end measurements from the clients.
Clients execute in a closed loop.
We use micro-benchmarks commonly used to evaluate BFT systems, e.g., {\small \sc BFT-SMaRt}.
Each experiment is run for $120$ seconds (including 30 s warm-up and cool-down).
The reported results are the average of five runs.

\subsection{Fault Tolerance and Scalability}

In the first set of experiments, we evaluate the throughput and latency of the protocols by increasing the number of replicas $n$ in a failure-free situation.
We vary the number of replicas in an experiment from $4$ to $100$. For some protocols, the smallest network size might differ, e.g., FaB requires $5f+1=6$ replicas.
We use batching with the batch size of $400$ (we discuss this choice in later experiments) and
a workload with client request/reply payload sizes of $128/128$ byte.
Figure~\ref{fig:nodes} reports the results.

Zyzzyva shows the highest throughput among all protocols in small networks due to its optimistic ordering stage
that does not require any communication among replicas (design choice \ref{tr:2phase}).
However, as $n$ increases, its throughput significantly reduces as clients need to wait for \reply from {\em all} replicas.
Increasing the number of replicas also has a large impact on PBFT and FaB ($65\%$ and $63\%$ reduction, respectively)
due to their quadratic message complexity.

On the other hand, the throughput of Kauri and \ftb is less affected ($31\%$ and $32\%$ reduction, respectively)
by increasing $n$ because of their tree topology (design choice \ref{tr:load}) that reduced the bandwidth utilization of each replica. 
Similarly, PoE, SBFT and HotStuff incur less throughput reduction ($39\%$, $55\%$ and $45\%$ respectively)
compared to PBFT and FaB due to their linear message complexity (design choice \ref{tr:phase}).
It should be noted that in \sys, chained HotStuff has been implemented using the pipelining technique
(optimization O \ref{optim:pipe}).
Hence, the average latency of requests has been reduced.
In comparison to HotStuff, SBFT has slightly lower throughput in large networks (e.g., $8\%$ lower when $n=100$)
because the leader waits for messages from all replicas.
SBFT, on the other hand, shows higher throughput compared to HotStuff in smaller networks (e.g., $12\%$  higher when $n=4$)
due to its fast ordering stage (design choice \ref{tr:1phase}).
PoE demonstrates higher throughput compared to both SBFT and HotStuff,
especially in larger networks (e.g., $39\%$ higher than SBFT and $26\%$ higher than HotStuff when $n=100$).
This is expected because, in PoE, the leader does not need to wait for messages from all replicas
and optimistically combines signatures from $2f+1$ replicas (design choice \ref{tr:s1phase}).
Compared to PBFT, while HotStuff demonstrates better throughput (e.g., $48\%$ higher when $n=64$),
the latency of PBFT is lower (e.g., $32\%$ lower when $n=64$).
One reason behind the high latency of HotStuff is its extra communication round (design choice \ref{tr:view1}).

\begin{figure}[t]
\centering
\begin{minipage}{.33\textwidth}\centering
\vspace{0.9em}
\begin{tikzpicture}[scale=0.7]
\begin{axis}[
    xlabel={$f$ value},
    ylabel={Throughput [ktrans/sec]},
    xmin=0, xmax=22,
    ymin=0, ymax=260,
    xtick={1,5,10,20},
    ytick={50,100,150,200,250},
    legend columns=10,
    legend style={at={(axis cs:1,-45)},anchor=north west}, 
    ymajorgrids=true,
    grid style=dashed,
]

\addplot[
    color=violet,
    mark=*,
    mark size=3pt,
    line width=0.2mm,
    ]
    coordinates {
    (1,148)(5,116)(10,95)(20,75)};

\addplot[
    color=cyan,
    mark=pentagon*,
    mark size=3pt,
    line width=0.2mm,
    ]
    coordinates {
    (1,250)(5,218)(10,191)(20,167)};
    
\addplot[
    color=magenta,
    mark=diamond,
    mark size=3pt,
    line width=0.2mm,
    ]
    coordinates {
    (1,190)(5,167)(10,143)(20,110)};
    
\addplot[
    color=orange,
    mark=diamond*,
    mark size=3pt,
    line width=0.2mm,
    ]
    coordinates {
    (1,194)(5,175)(10,154)(20,140)};
    
\addplot[
    color=black,
    mark=pentagon,
    mark size=3pt,
    line width=0.2mm,
    ]
    coordinates {
    (1,175)(5,129)(10,104)(20,65)};
    
\addplot[
    color=purple,
    mark=o,
    mark size=3pt,
    line width=0.2mm,
    ]
    coordinates {
    (1,170)(5,154)(10,131)(20,116)};
    
 \addplot[
    color=red,
    mark=square,
    mark size=3pt,
    line width=0.2mm,
    ]
    coordinates {
    (1,142)(5,121)(10,109)(20,100)};

\addplot[
    color=blue,
    mark=triangle*,
    mark size=3pt,
    line width=0.2mm,
    ]
    coordinates {
    (1,28)(5,22)(10,18)(20,15)};
    
 \addplot[
    color=teal,
    mark=otimes,
    mark size=3pt,
    line width=0.2mm,
    ]
    coordinates {
    (1,194)(5,174)(10,148)(20,115)};
    
 \addplot[
    color=gray,
    mark=square*,
    mark size=3pt,
    line width=0.2mm,
    ]
    coordinates {
    (1,177)(5,167)(10,155)(20,122)};

\addlegendentry{PBFT}
\addlegendentry{Zyzzyva}
\addlegendentry{SBFT}
\addlegendentry{PoE}
\addlegendentry{FaB}
\addlegendentry{HotStuff}
\addlegendentry{Kuari}
\addlegendentry{Themis}
\addlegendentry{\flb}
\addlegendentry{\ftb}
 
\end{axis}
\end{tikzpicture}
\end{minipage}\hspace{4em}
\begin{minipage}{.33\textwidth} \centering
\begin{tikzpicture}[scale=0.7]
\begin{axis}[
    xlabel={$f$ value},
    ylabel={Latency [ms]},
    xmin=0, xmax=21,
    ymin=0, ymax=100,
    xtick={1,5,10,20},
    ytick={0,20,40,60,80}, 
    ymajorgrids=true,
    grid style=dashed,
]

\addplot[
    color=violet,
    mark=*,
    mark size=3pt,
    line width=0.2mm,
    ]
    coordinates {
    (1,3)(5,7)(10,11)(20,27)};

\addplot[
    color=cyan,
    mark=pentagon*,
    mark size=3pt,
    line width=0.2mm,
    ]
    coordinates {
    (1,3)(5,4)(10,10)(20,23)};
    
\addplot[
    color=magenta,
    mark=diamond,
    mark size=3pt,
    line width=0.2mm,
    ]
    coordinates {
    (1,3)(5,6)(10,10)(20,24)};
    
\addplot[
    color=orange,
    mark=diamond*,
    mark size=3pt,
    line width=0.2mm,
    ]
    coordinates {
    (1,3)(5,6)(10,9)(20,15)};
    
\addplot[
    color=black,
    mark=pentagon,
    mark size=3pt,
    line width=0.2mm,
    ]
    coordinates {
    (1,3)(5,7)(10,15)(20,38)};
    
\addplot[
    color=purple,
    mark=o,
    mark size=3pt,
    line width=0.2mm,
    ]
    coordinates {
    (1,5)(5,11)(10,17)(20,35)};
    
 \addplot[
    color=red,
    mark=square,
    mark size=3pt,
    line width=0.2mm,
    ]
    coordinates {
    (1,8)(5,18)(10,31)(20,58)};

\addplot[
    color=blue,
    mark=triangle*,
    mark size=3pt,
    line width=0.2mm,
    ]
    coordinates {
    (1,13)(5,21)(10,38)(20,89)};
    
 \addplot[
    color=teal,
    mark=otimes,
    mark size=3pt,
    line width=0.2mm,
    ]
    coordinates {
    (1,3)(5,9)(10,15)(20,31)};
    
 \addplot[
    color=gray,
    mark=square*,
    mark size=3pt,
    line width=0.2mm,
    ]
    coordinates {
    (1,6)(5,18)(10,31)(20,63)};

\end{axis}
\end{tikzpicture}
\end{minipage}
\caption{Performance with different $f$ value}
  \label{fig:failures}
\end{figure}
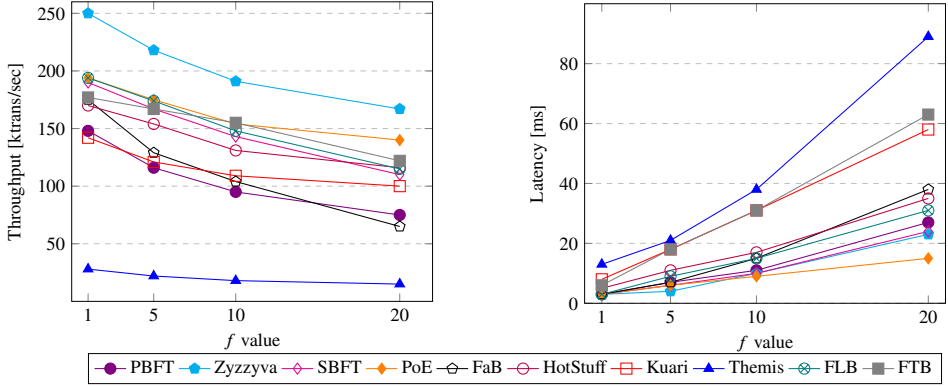

Supporting order-fairness (design choice \ref{tr:fair}) leads to
deficient performance of Themis compared to HotStuff ($83\%$ lower throughput when $n=5$).
In Themis, replicas need to order transactions and send batches of transactions to the leader,
and the leader needs to generate a fair order.
As the number of replicas increases, Themis incurs higher latency
(the latency increases from $9$ ms to $137$ as the $n$ increases from $5$ to $101$).
One main source of latency is the time the leader takes to generate the dependency graph and reach a fair order.
It should also be noted that in the \sys implementation of Themis, ZKP has not been used and the leader sends
all transaction orderings received from replicas to all of them in the \two phase. This might slightly increase the latency.
Using design choice \ref{tr:nodes} and reducing the number of communication phases results in
$41\%$ higher throughput and $46\%$ lower latency of \ftb compared to Kauri in a setting with $99$ replicas ($100$ for Kauri).

Finally, in \flb, by combining design choices \ref{tr:phase} and \ref{tr:nodes} demonstrates high throughput and low latency
for large value of $n$ ($2.25$x throughput and $0.55$x latency compared to PBFT). 
This is expected because first, \flb reduces both message complexity and communication phases, and
second, in contrast to SBFT and Zyzzyva, replicas in \flb do not need to wait for responses from all other replicas.

Figure~\ref{fig:nodes} demonstrates the performance of protocols with different numbers of replicas.
However, with the same number of replicas, different protocols tolerate different numbers of failures.
For instance, PBFT requires $3f+1$ and when $n=100$ tolerates $33$ failures while FaB requires $5f+1$ and tolerates $19$ failures with $n=100$.
To compare protocols based on the maximum number of tolerated failures,
we represent the results of the first experiments in Figure~\ref{fig:failures}.
When protocols tolerate $20$ failures, Themis incurs the highest latency.
This is because Themis requires $81$ ($4f+1$) replicas and deals with the high cost of achieving order-fairness.

\subsection{Performance with Faulty Backups}

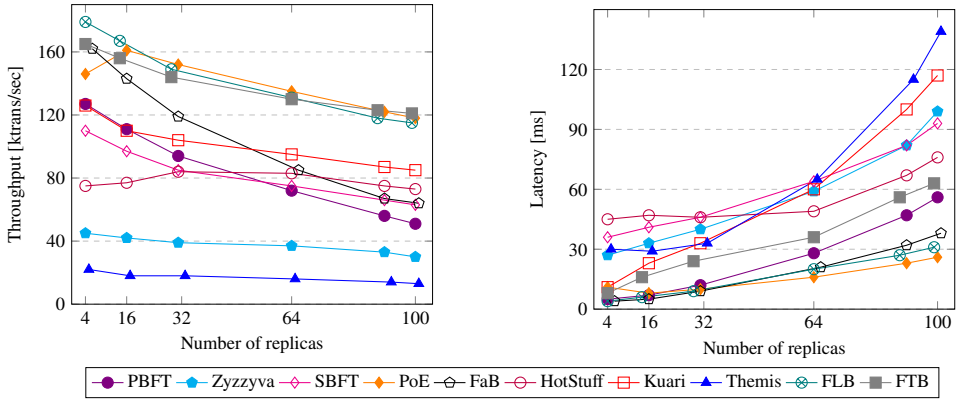
\begin{figure}[t]
\centering
\begin{minipage}{.33\textwidth}\centering
\vspace{1.1em}
\begin{tikzpicture}[scale=0.7]
\begin{axis}[
    xlabel={Number of replicas},
    ylabel={Throughput [ktrans/sec]},
    xmin=0, xmax=105,
    ymin=0, ymax=190,
    xtick={4,16,32,64,100},
    ytick={0,40,80,120,160},
    legend columns=10,
    legend style={at={(axis cs:4,-40)},anchor=north west}, 
    ymajorgrids=true,
    grid style=dashed,
]

\addplot[
    color=violet,
    mark=*,
    mark size=3pt,
    line width=0.2mm,
    ]
    coordinates {
    (4,127)(16,111)(31,94)(64,72)(91,56)(100,51)};

\addplot[
    color=cyan,
    mark=pentagon*,
    mark size=3pt,
    line width=0.2mm,
    ]
    coordinates {
    (4,45)(16,42)(31,39)(64,37)(91,33)(100,30)};

\addplot[
    color=magenta,
    mark=diamond,
    mark size=3pt,
    line width=0.2mm,
    ]
    coordinates {
    (4,110)(16,97)(31,85)(64,75)(91,66)(100,63)};
    
\addplot[
    color=orange,
    mark=diamond*,
    mark size=3pt,
    line width=0.2mm,
    ]
    coordinates {
    (4,146)(16,161)(31,152)(64,135)(91,122)(100,118)};
    
\addplot[
    color=black,
    mark=pentagon,
    mark size=3pt,
    line width=0.2mm,
    ]
    coordinates {
    (6,162)(16,143)(31,119)(66,85)(91,67)(101,64)};
    
\addplot[
    color=purple,
    mark=o,
    mark size=3pt,
    line width=0.2mm,
    ]
    coordinates {
    (4,75)(16,77)(31,84)(64,83)(91,75)(100,73)};

 \addplot[
    color=red,
    mark=square,
    mark size=3pt,
    line width=0.2mm,
    ]
    coordinates {
    (4,126)(16,110)(31,104)(64,95)(91,87)(100,85)};

\addplot[
    color=blue,
    mark=triangle*,
    mark size=3pt,
    line width=0.2mm,
    ]
    coordinates {
    (5,22)(17,18)(33,18)(65,16)(93,14)(101,13)};

 \addplot[
    color=teal,
    mark=otimes,
    mark size=3pt,
    line width=0.2mm,
    ]
    coordinates {
    (4,179)(14,167)(29,149)(64,131)(89,118)(99,115)};

 \addplot[
    color=gray,
    mark=square*,
    mark size=3pt,
    line width=0.2mm,
    ]
    coordinates {
    (4,165)(14,156)(29,144)(64,130)(89,123)(99,121)};

\addlegendentry{PBFT}
\addlegendentry{Zyzzyva}
\addlegendentry{SBFT}
\addlegendentry{PoE}
\addlegendentry{FaB}
\addlegendentry{HotStuff}
\addlegendentry{Kuari}
\addlegendentry{Themis}
\addlegendentry{\flb}
\addlegendentry{\ftb}
 
\end{axis}
\end{tikzpicture}
\end{minipage}\hspace{4em}
\begin{minipage}{.33\textwidth} \centering
\begin{tikzpicture}[scale=0.7]
\begin{axis}[
    xlabel={Number of replicas},
    ylabel={Latency [ms]},
    xmin=0, xmax=105,
    ymin=0, ymax=150,
    xtick={4,16,32,64,100},
    ytick={0,30,60,90,120}, 
    ymajorgrids=true,
    grid style=dashed,
]

\addplot[
    color=violet,
    mark=*,
    mark size=3pt,
    line width=0.2mm,
    ]
    coordinates {
    (4,5)(16,7)(31,12)(64,28)(91,47)(100,56)};

\addplot[
    color=cyan,
    mark=pentagon*,
    mark size=3pt,
    line width=0.2mm,
    ]
    coordinates {
    (4,27)(16,33)(31,40)(64,59)(91,82)(100,99)};

\addplot[
    color=magenta,
    mark=diamond,
    mark size=3pt,
    line width=0.2mm,
    ]
    coordinates {
    (4,36)(16,41)(31,46)(64,64)(91,82)(100,93)};
    
\addplot[
    color=orange,
    mark=diamond*,
    mark size=3pt,
    line width=0.2mm,
    ]
    coordinates {
    (4,11)(16,8)(31,10)(64,16)(91,23)(100,26)};
    
\addplot[
    color=black,
    mark=pentagon,
    mark size=3pt,
    line width=0.2mm,
    ]
    coordinates {
    (6,4)(16,5)(31,9)(66,21)(91,32)(101,38)};
    
\addplot[
    color=purple,
    mark=o,
    mark size=3pt,
    line width=0.2mm,
    ]
    coordinates {
    (4,45)(16,47)(31,46)(64,49)(91,67)(100,76)};
    
 \addplot[
    color=red,
    mark=square,
    mark size=3pt,
    line width=0.2mm,
    ]
    coordinates {
    (4,11)(16,23)(31,33)(64,60)(91,100)(100,117)};

\addplot[
    color=blue,
    mark=triangle*,
    mark size=3pt,
    line width=0.2mm,
    ]
    coordinates {
    (5,30)(17,29)(33,33)(65,65)(93,115)(101,139)};
    
 \addplot[
    color=teal,
    mark=otimes,
    mark size=3pt,
    line width=0.2mm,
    ]
    coordinates {
    (4,4)(14,6)(29,9)(64,20)(89,27)(99,31)};
    
 \addplot[
    color=gray,
    mark=square*,
    mark size=3pt,
    line width=0.2mm,
    ]
    coordinates {
    (4,8)(14,16)(29,24)(64,36)(89,56)(99,63)};

\end{axis}
\end{tikzpicture}
\end{minipage}
\caption{Performance with faulty backups}
  \label{fig:faulty}
\end{figure}

In this set of experiments, we force a backup replica to fail and repeat the first set of experiments.
Figure~\ref{fig:faulty} reports the results.
Zyzzyva is mostly affected by failures ($82\%$ lower throughput) as clients need to collect
responses from {\em all} replicas.
A client waits for $\Delta = 5 ms$ to receive \reply from all replicas and then
the protocol switches to its normal path).

We also run this experiment with and without faulty backups on Zyzzyva5
to validate design choice \ref{tr:resilience},
i.e., tolerating $f$ faulty replicas by increasing the number of replicas.
With a single faulty backup, Zyzzyva5 incurs only $8\%$ lower throughput when $n=6$.

Backup failure reduces the throughput of SBFT by $42\%$.
In the fast path of SBFT, all replicas need to participate, and even when a single replica is faulty, the protocol
falls back to its slow path that requires two more phases.
Interestingly, while the throughput of PoE is reduced by $26\%$ in a small network ($4$ replicas), its throughput
is not significantly affected in large networks.
This is because the chance of the faulty replica
(which participates in the quorum construction but does not send \reply messages to the clients)
to be a member of the quorum is much higher in small networks.

Faulty backups also affect the performance of HotStuff, especially in small networks.
This is expected because HotStuff uses the rotating leader mechanism.
When $n$ is small, the faulty replica is the leader of more views during the experiments, resulting in reduced performance.
HotStuff demonstrates its best performance when $n=31$
(still, $36\%$ lower throughput and $2.7$x latency compared to the failure-free scenario).
While Themis uses HotStuff as its ordering stage, a single faulty backup has less impact on its performance
compared to HotStuff ($25\%$ reduction vs. $66\%$ reduction in throughput).
This is because Themis has a larger network size ($4f+1$ vs. $3f+1$) that reduces the impact of the faulty replica.
In Kauri and \ftb, we force a leaf replica to fail in order to avoid triggering a reconfiguration. As a result,
the failure of a backup does not significantly affect their performance (e.g., $3\%$ lower throughput with $31$ replicas in Kauri).
Finally, in small networks, \flb demonstrates the best performance as it incurs only $8\%$ throughput reduction.

\subsection{Impact of Request Batching}

In the next set of experiments, we measure the impact of request batching on the performance of different protocols
implemented in \sys.
We consider three scenarios with batch size of $200$, $400$ and $800$.
The network includes $16$ replicas ($17$ replicas for Themis, $14$ replicas for \flb and \ftb) and all replicas are non-faulty.
Figure~\ref{fig:batching} depicts the results for three batch sizes of $200$, $400$ and $800$.

Increasing the batch size from $200$ to $400$ requests improves the performance of all protocols.
This is expected because, with larger batch sizes, more transactions can be committed while
the number of communication phases and exchanged messages is the same and
the bandwidth and computing resources are not fully utilized yet.
Different protocols behave differently when the batch size increases from $400$ to $800$.
First, Kauri and \ftb still process a higher number of transactions ($42\%$ and $34\%$ higher throughput).
This is because Kauri and \ftb balance the load and utilize the bandwidth of all replicas.
Second, SBFT and FaB demonstrate similar performance as before;
a trade-off between smaller consensus quorums and higher cost of signature verification and bandwidth utilization.
Third, the performance of Themis decreases ($24\%$ lower throughput and $3.16$x latency)
compared to a batch size of $400$ due to two main reasons.
First, the higher cost of signature verification and bandwidth utilization, and
second, the higher complexity of generating fair order for a block of $800$ transactions (CPU utilization).

\definecolor{violet_new}{RGB}{218, 153, 247}
\definecolor{gray_new}{RGB}{218, 213, 219}
\definecolor{blue_new}{RGB}{155, 198, 224}
\definecolor{purple_new}{RGB}{245, 169, 195}

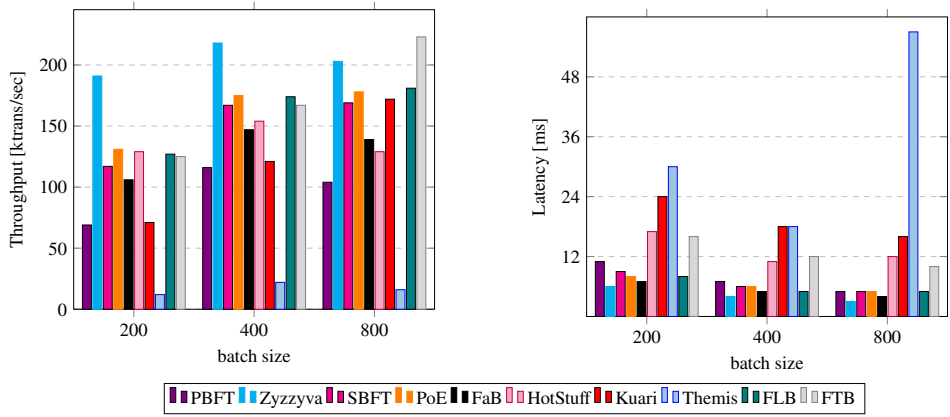
\begin{figure}[t]
\centering
\begin{minipage}{.33\textwidth} \centering
\vspace{1.2em}
\begin{tikzpicture}[scale=0.7]
    \begin{axis}[
       ybar=2*\pgflinewidth,
        bar width=0.17cm,
        ymajorgrids = true,
        grid style=dashed,
        xlabel={batch size},
        ylabel={Throughput [ktrans/sec]},
        symbolic x coords={200,400,800},
        xtick = data,
        scaled y ticks = false,
        enlarge x limits=0.25,
        ymin=0,
        legend columns=10,
        legend style={at={(0.25,-0.35)},anchor=south west},
    ]
    
    \addplot[style={black,fill=violet}]
    coordinates {(200,69)(400,116)(800,104)};
 
    \addplot[style={cyan,fill=cyan,mark=none}]
    coordinates {(200,191)(400,218)(800,203)};

    \addplot[style={black,fill=magenta}]
    coordinates {(200,117)(400,167)(800,169)};
    
    \addplot[style={orange,fill=orange,mark=none}]
    coordinates {(200,131)(400,175)(800,178)};
    
    \addplot[style={black,fill=black}]
    coordinates {(200,106)(400,147)(800,139)};
    
    \addplot[style={purple,fill=purple_new,mark=none}]
    coordinates {(200,129)(400,154)(800,129)};
    
    \addplot[style={black,fill=red}]
    coordinates {(200,71)(400,121)(800,172)};
    
    \addplot[style={blue,fill=blue_new,mark=none}]
    coordinates {(200,12)(400,22)(800,16)};
    
    \addplot[style={black,fill=teal}]
    coordinates {(200,127)(400,174)(800,181)};
    
    \addplot[style={gray,fill=gray_new,mark=none}]
    coordinates {(200,125)(400,167)(800,223)};

\addlegendentry{PBFT}
\addlegendentry{Zyzzyva}
\addlegendentry{SBFT}
\addlegendentry{PoE}
\addlegendentry{FaB}
\addlegendentry{HotStuff}
\addlegendentry{Kuari}
\addlegendentry{Themis}
\addlegendentry{\flb}
\addlegendentry{\ftb}

    \end{axis}
\end{tikzpicture}
\end{minipage}\hspace{4em}
\begin{minipage}{.33\textwidth} \centering
\begin{tikzpicture}[scale=0.7]
    \begin{axis}[
       ybar=2*\pgflinewidth,
        bar width=0.17cm,
        ymajorgrids = true,
        grid style=dashed,
        xlabel={batch size},
        ylabel={Latency [ms]},
        symbolic x coords={200,400,800},
        xtick = data,
        ymax=60,
        ytick={12,24,36,48},
        scaled y ticks = false,
        enlarge x limits=0.25,
        ymin=0, 
    ]
    
    \addplot[style={black,fill=violet}]
    coordinates {(200,11)(400,7)(800,5)};
 
    \addplot[style={cyan,fill=cyan,mark=none}]
    coordinates {(200,6)(400,4)(800,3)};

    \addplot[style={black,fill=magenta}]
    coordinates {(200,9)(400,6)(800,5)};
    
    \addplot[style={orange,fill=orange,mark=none}]
    coordinates {(200,8)(400,6)(800,5)};
    
    \addplot[style={black,fill=black}]
    coordinates {(200,7)(400,5)(800,4)};
    
    \addplot[style={purple,fill=purple_new,mark=none}]
    coordinates {(200,17)(400,11)(800,12)};
    
    \addplot[style={black,fill=red}]
    coordinates {(200,24)(400,18)(800,16)};
    
    \addplot[style={blue,fill=blue_new,mark=none}]
    coordinates {(200,30)(400,18)(800,57)};
    
    \addplot[style={black,fill=teal}]
    coordinates {(200,8)(400,5)(800,5)};
    
    \addplot[style={gray,fill=gray_new,mark=none}]
    coordinates {(200,16)(400,12)(800,10)};

    \end{axis}
\end{tikzpicture}
\end{minipage}
\caption{Impact of request batching}
  \label{fig:batching}
\end{figure}

\subsection{Impact of a Geo-distributed Setup}

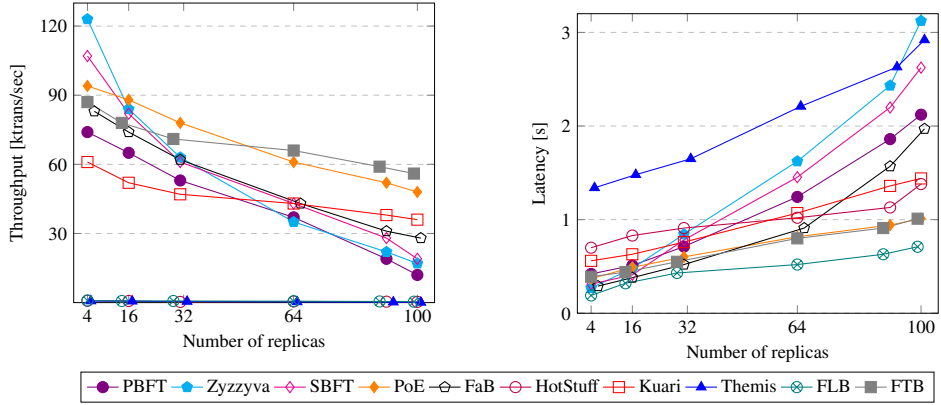
\begin{figure}[t]
\centering
\begin{minipage}{.33\textwidth}\centering
\vspace{0.9em}
\begin{tikzpicture}[scale=0.7]
\begin{axis}[
    xlabel={Number of replicas},
    ylabel={Throughput [ktrans/sec]},
    xmin=0, xmax=105,
    ymin=0, ymax=130,
    xtick={4,16,32,64,100},
    ytick={30,60,90,120},
    legend columns=10,
    legend style={at={(axis cs:2,-30)},anchor=north west}, 
    ymajorgrids=true,
    grid style=dashed,
]

\addplot[
    color=violet,
    mark=*,
    mark size=3pt,
    line width=0.2mm,
    ]
    coordinates {
    (4,74)(16,65)(31,53)(64,37)(91,19)(100,12)};

\addplot[
    color=cyan,
    mark=pentagon*,
    mark size=3pt,
    line width=0.2mm,
    ]
    coordinates {
    (4,123)(16,84)(31,63)(64,35)(91,22)(100,17)};

\addplot[
    color=magenta,
    mark=diamond,
    mark size=3pt,
    line width=0.2mm,
    ]
    coordinates {
    (4,107)(16,82)(31,61)(64,43)(91,28)(100,19)};
    
\addplot[
    color=orange,
    mark=diamond*,
    mark size=3pt,
    line width=0.2mm,
    ]
    coordinates {
    (4,94)(16,88)(31,78)(64,61)(91,52)(100,48)};
    
\addplot[
    color=black,
    mark=pentagon,
    mark size=3pt,
    line width=0.2mm,
    ]
    coordinates {
    (6,83)(16,74)(31,62)(66,43)(91,31)(101,28)}; 
    
\addplot[
    color=purple,
    mark=o,
    mark size=3pt,
    line width=0.2mm,
    ]
    coordinates {
    (4,0.8)(16,0.7)(31,0.3)(64,0.5)(91,0.4)(100,0.3)};
    
 \addplot[
    color=red,
    mark=square,
    mark size=3pt,
    line width=0.2mm,
    ]
    coordinates {
    (4,61)(16,52)(31,47)(64,43)(91,38)(100,36)};

\addplot[
    color=blue,
    mark=triangle*,
    mark size=3pt,
    line width=0.2mm,
    ]
    coordinates {
    (5,0.7)(17,0.6)(33,0.5)(65,0.3)(93,0.2)(101,0.1)};
    
 \addplot[
    color=teal,
    mark=otimes,
    mark size=3pt,
    line width=0.2mm,
    ]
    coordinates {
    (4,0.9)(14,0.7)(29,0.7)(64,0.6)(89,0.5)(99,0.4)};
    
 \addplot[
    color=gray,
    mark=square*,
    mark size=3pt,
    line width=0.2mm,
    ]
    coordinates {
    (4,87)(14,78)(29,71)(64,66)(89,59)(99,56)};

\addlegendentry{PBFT}
\addlegendentry{Zyzzyva}
\addlegendentry{SBFT}
\addlegendentry{PoE}
\addlegendentry{FaB}
\addlegendentry{HotStuff}
\addlegendentry{Kuari}
\addlegendentry{Themis}
\addlegendentry{\flb}
\addlegendentry{\ftb}
 
\end{axis}
\end{tikzpicture}
\end{minipage}\hspace{4em}
\begin{minipage}{.33\textwidth} \centering
\begin{tikzpicture}[scale=0.7]
\begin{axis}[
    xlabel={Number of replicas},
    ylabel={Latency [s]},
    xmin=0, xmax=105,
    ymin=0, ymax=3.2,
    xtick={4,16,32,64,100},
    ytick={0,1,2,3}, 
    ymajorgrids=true,
    grid style=dashed,
]

\addplot[
    color=violet,
    mark=*,
    mark size=3pt,
    line width=0.2mm,
    ]
    coordinates {
    (4,0.42)(16,0.51)(31,0.712)(64,1.243)(91,1.86)(100,2.121)};

\addplot[
    color=cyan,
    mark=pentagon*,
    mark size=3pt,
    line width=0.2mm,
    ]
    coordinates {
    (4,0.27)(16,0.48)(31,0.843)(64,1.623)(91,2.432)(100,3.123)};
    
\addplot[
    color=magenta,
    mark=diamond,
    mark size=3pt,
    line width=0.2mm,
    ]
    coordinates {
    (4,0.31)(16,0.41)(31,0.783)(64,1.453)(91,2.198)(100,2.623)};

\addplot[
    color=orange,
    mark=diamond*,
    mark size=3pt,
    line width=0.2mm,
    ]
    coordinates {
    (4,0.37)(16,0.49)(31,0.602)(64,0.817)(91,0.942)(100,1.011)};

\addplot[
    color=black,
    mark=pentagon,
    mark size=3pt,
    line width=0.2mm,
    ]
    coordinates {
    (6,0.29)(16,0.38)(31,0.52)(66,0.91)(91,1.57)(101,1.97)};
    
\addplot[
    color=purple,
    mark=o,
    mark size=3pt,
    line width=0.2mm,
    ]
    coordinates {
    (4,0.7)(16,0.83)(31,0.91)(64,1.02)(91,1.13)(100,1.38)};
    
 \addplot[
    color=red,
    mark=square,
    mark size=3pt,
    line width=0.2mm,
    ]
    coordinates {
    (4,0.56)(16,0.63)(31,0.76)(64,1.07)(91,1.36)(100,1.44)};

\addplot[
    color=blue,
    mark=triangle*,
    mark size=3pt,
    line width=0.2mm,
    ]
    coordinates {
    (5,1.34)(17,1.48)(33,1.65)(65,2.21)(93,2.63)(101,2.92)};
    
 \addplot[
    color=teal,
    mark=otimes,
    mark size=3pt,
    line width=0.2mm,
    ]
    coordinates {
    (4,0.19)(14,0.32)(29,0.43)(64,0.52)(89,0.63)(99,0.71)};
    
 \addplot[
    color=gray,
    mark=square*,
    mark size=3pt,
    line width=0.2mm,
    ]
    coordinates {
    (4,0.39)(14,0.44)(29,0.55)(64,0.80)(89,0.91)(99,1.01)};

\end{axis}
\end{tikzpicture}
\end{minipage}
\caption{Performance with a geo-distributed setup}
  \label{fig:distribution}
\end{figure}

We measure the performance of different protocols in a wide-area network.
Replicas are deployed in 4 different AWS regions, i.e.,
Tokyo ({\em TY}), Seoul ({\em SU}), Virginia ({\em VA}), and California ({\em CA}) with
an average Round-Trip Time (RTT) of
{\em TY} $\rightleftharpoons$ {\em SU}: $33$ ms,
{\em TY} $\rightleftharpoons$ {\em VA}: $148$ ms,
{\em TY} $\rightleftharpoons$ {\em CA}: $107$ ms,
{\em SU} $\rightleftharpoons$ {\em VA}: $175$ ms,
{\em SU} $\rightleftharpoons$ {\em CA}: $135$ ms, and
{\em VA} $\rightleftharpoons$ {\em CA}: $62$ ms.
The clients are also placed in Oregon ({\em OR}) with an average RTT of
$97$, $126$, $68$ and $22$ ms from {\em TY}, {\em SU}, {\em VA} and {\em CA} respectively.
We use a batch size of $400$ and perform experiments in a failure-free situation.
In this experiment, the pipelining stretch of Kauri and \ftb  is increased to $6$.
Figure~\ref{fig:distribution} depicts the results.

Zyzzyva demonstrates the best performance when $n$ is small. However, when $n$ increases,
its performance is significantly reduced ($87\%$ throughput reduction and $115$x latency when $n$ increases from $4$ to $100$).
This is because, in Zyzzyva, clients need to receive \reply messages from {\em all} replicas.
Similarly, SBFT incurs a significant reduction in its performance due to its optimistic assumption that all replicas participate in a timely manner.
In both protocols, replicas (client or leader) wait for $\Delta = 500$ ms to receive responses from all replicas before switching to the normal path.
This reduction can be seen in PBFT as well ($84\%$ throughput reduction when $n$ increases to $100$) due to its
quadratic communication complexity.
PoE incurs a smaller throughput reduction ($51\%$) in comparison to Zyzzyva, SBFT, and PBFT because it does not need
to wait for all replicas and it has a linear communication complexity.
Increasing the number of replicas does not significantly affect the throughput of \ftb compared to other protocols
($36\%$ throughput reduction when $n$ increases to $99$)
due to its logarithmic message complexity and pipelining technique.

In HotStuff, the leader of the following view must wait for the previous view's decision before initiating its value.
Even though Chained HotStuff is implemented in \sys, the leader still needs to wait for one communication round (an RTT).
As a result, in contrast to the single datacenter setting where each round takes ${\sim} 1$ ms,
request batches are proposed on average every ${\sim} 190$ ms.
In this setting, a larger batch size possibly improves the performance of HotStuff.
Similarly, in Themis and \flb, the leader must wait for certificates from $n-f$ replicas before initiating consensus on the next request batch.
In Themis, network latency also affects achieving order-fairness as replicas might propose different orders for client requests.
This result demonstrates the significant impact of the out-of-order processing of requests
(optimization O\ref{optim:order})
on the performance of the protocol, especially in a wide area network. 

\subsection{Evaluation Summary}
We summarize some of the evaluation results as follows.
First, optimistic protocols that require all nodes to participate, e.g., Zyzzyva and SBFT,
do not perform well in large networks, especially when nodes are far apart.
In small networks also, a single faulty node significantly reduces the performance of optimistic protocols.
Second, the performance of pessimistic protocols highly depends on the communication topology.
While the performance of protocols with quadratic communication complexity, e.g., PBFT and FaB, is
significantly reduced by increasing the network size, the performance of protocols with linear complexity, e.g., HotStuff, and
especially logarithmic complexity, e.g., Kauri and \ftb, is less affected.
Interestingly in small networks, protocols that use the leader rotation mechanism show poor performance. This is because the chance of the faulty node becoming the leader is relatively high.
Third, increasing the request batch size enhances the performance of all protocols
to the point where the bandwidth and computing resources are fully utilized.
However, the load-balancing techniques, e.g., tree topology, enable a protocol to process larger batches.
Finally, in a wide area network, out-of-order processing of transactions significantly improves performance.
In such a setting, protocols that require a certificate of the previous round to start a new round, e.g., HotStuff,
show poor performance even with pipelining techniques.
\section{Related Work}\label{sec:related}

SMR regulates the
deterministic execution of client requests on
multiple replicas, such that
every non-faulty replica executes every request in 
the same order \cite{schneider1990implementing,lamport1978time}.
Several approaches \cite{schneider1990implementing,lamport2001paxos,ongaro2014search}
generalize SMR to support crash failures.
CFT protocols \cite{lamport2019part,van2014vive,ongaro2014search,lamport2005generalized,oki1988viewstamped,li2007paxos,lampson2001abcd,lamport2006fast,chandra2007paxos,brasileiro2001consensus,junqueira2011zab,lamport2004cheap,howard2017flexible,ailijiang2019wpaxos,liskov2012viewstamped,nawab2018dpaxos,howard2017flexible,charapko2021pigpaxos,dobre2010hp,pedone2001boosting}
utilize the design trade-offs between different design dimensions.
For instance, Fast Paxos \cite{lamport2006fast} adds $f$ more replicas to reduce one phase of communication.

Byzantine fault tolerance refers to nodes that behave arbitrarily
after the seminal work by Lamport, et al. \cite{lamport1982byzantine}.
BFT protocols have been analyzed in several surveys and empirical studies \cite{singh2008bft,cachin2017blockchain,abraham2017blockchain,alqahtani2021bottlenecks,gupta2016bft,gai2021dissecting,bessani2014state,amiri2020modern,correia2011byzantine,berger2018scaling,platania2016choosing,distler2021byzantine,bano2019sok,zhang2022reaching,abraham2017revisiting}.
We discuss some of the more relevant studies.

Berger and Reiser~\cite{berger2018scaling} present
a survey on BFT protocols used in blockchains
where the focus is on the scalability techniques.
Similarly, a survey on BFT protocols consisting of
classical protocols, e.g., PBFT, blockchain protocols, e.g., PoW,
and hybrid protocols, e.g., OmniLedger \cite{kokoris2018omniledger},
and their applications in permissionless blockchains, is conducted by Bano et al.~\cite{bano2019sok}.
Platania et al.~\cite{platania2016choosing} classify BFT protocols into client-side and server-side protocols
depending on the client's role. The paper compares these two classes of protocols and
analyzes their performance and correctness attacks.
Three families of leader-based, leaderless, and robust BFT protocols
with a focus on message and time complexities have been analyzed by Zhang et al.~\cite{zhang2022reaching}.
Finally, Distler~\cite{distler2021byzantine} analyzes BFT protocols along
several main dimensions: architecture, clients, agreement, execution, checkpoint, and recovery.
The paper shares several dimensions with \sys.

A recent line of work \cite{abraham2020brief, abraham2021good, abraham2020sync,abraham2021goodun} also study
good-case latency of BFT protocols.
\sys, in contrast to all these survey and analysis papers,
provides a design space, systematically discusses design choices (trade-offs),
and, more importantly, provides a tool to experimentally analyze BFT protocols.

BFTSim \cite{singh2008bft} is a simulation environment for BFT protocols that leverages declarative networking system.
The paper also compares a set of representative protocols using the simulator.
Abstract \cite{aublin2015next} presents a framework to design and reconfigure BFT protocols where
each protocol is developed as a sequence of BFT instances.
Abstract presents AZyzzyva, Aliph, and R-Aliph as three BFT protocols.
Each protocol itself is a composition of Abstract instances
presented to handle different situations (e.g., fault-free, under attack).
For instance, R-Aliph is a composition of Quorum, Chain, and Aardvark.
In contrast to such studies, \sys attempts to develop a unified design space for BFT protocols, enabling
end-users to choose a protocol that best fits their application requirements.

In addition to CFT and BFT protocols, consensus with multiple failure modes has also been studied
for both
synchronous \cite{thambidurai1988interactive,meyer1991consensus,kieckhafer1994reaching,siu1996note}, and
partial synchronous \cite{porto2015visigoth,liu2016xft,gueta2019sbft,serafini2010scrooge,clement2009upright,amiri2020seemore}
models.
Finally, leaderless protocols \cite{lamport2011brief,duan2018beat,crain2018dbft,miller2016honey,suri2021basil,borran2010leader,guo2020dumbo}
have been proposed to avoid the implications of relying on a leader.
\section{Conclusion}\label{sec:conc}

We present \sys, a toolkit that unifies all BFT protocols within a platform
for analysis, design, implementation, and experimentation.
In using \sys, we demonstrate how different BFT protocols relate to one another within a design space
and evaluate BFT protocols under a unified deployment and experimentation environment in a fair and efficient manner.
By providing a unified platform for all the different BFT protocols,
\sys is able to highlight the strengths and weaknesses of diverse properties, e.g., optimistic vs. pessimistic.
The tool also provides a basis for discovering protocols not previously proposed.

While this paper focuses on the platform, the ultimate decision process
lies with the end-user for selecting and generating the BFT protocol.
As future work, we plan to explore incorporating automatic selection strategies in \sys
based on deployment environment and application requirements.
Machine learning techniques may be useful here in aiding the user in selecting the appropriate BFT protocol,
or evolving one protocol to another at runtime as system parameters are updated.
Furthermore, we will extend \sys by enabling protocol designers to define new dimensions and values systematically.

\balance
\bibliographystyle{abbrv}
\bibliography{_blockchain,_misc,_privacy,_system}
\end{document}